\documentclass[fleqn,usenatbib]{mnras}

\usepackage{newtxtext,newtxmath}
\usepackage{float}
\usepackage{pdflscape}
\usepackage{rotating}
\usepackage{multirow}
\usepackage{graphicx}
\usepackage{subcaption}
\restylefloat{table}

\usepackage[T1]{fontenc}
\usepackage{ae,aecompl}

\usepackage{graphicx}	
\usepackage{amsmath}	
\usepackage{amssymb}	

\title[STRIDES II]{The STRong lensing Insights into the Dark Energy Survey
	(STRIDES) 2016 follow-up campaign. II. New quasar
	lenses from double component fitting.}

\author[T. Anguita et al.]{
	\parbox{\textwidth}{
		\Large
		T. Anguita$^{1,2}$\thanks{E-mail: tanguita@gmail.com},
P.~L.~Schechter$^{3}$,
N.~Kuropatkin$^{4}$,
N.~D.~Morgan$^{5}$,
F.~Ostrovski$^{6,7}$,
L.~E.~Abramson$^{8}$,
A.~Agnello$^{9}$,
Y.~Apostolovski$^{1}$,
C.~D.~Fassnacht$^{10}$,
J.~W.~Hsueh$^{10}$,
V.~Motta$^{11}$,
K.~Rojas$^{11}$,
C.~E.~Rusu$^{12}$,
T.~Treu$^{8}$,
P.~Williams$^{8}$,
M.~Auger$^{6}$,
E.~Buckley-Geer$^{4}$,
H.~Lin$^{4}$,
R.~McMahon$^{6}$,
T.~M.~C.~Abbott$^{13}$,
S.~Allam$^{4}$,
J.~Annis$^{4}$,
R.~A.~Bernstein$^{14}$,
E.~Bertin$^{15,16}$,
D.~Brooks$^{17}$,
D.~L.~Burke$^{18,19}$,
A.~Carnero~Rosell$^{20,21}$,
M.~Carrasco~Kind$^{22,23}$,
J.~Carretero$^{24}$,
C.~E.~Cunha$^{18}$,
C.~B.~D'Andrea$^{25}$,
J.~De~Vicente$^{26}$,
D.~L.~DePoy$^{27}$,
S.~Desai$^{28}$,
H.~T.~Diehl$^{4}$,
P.~Doel$^{17}$,
B.~Flaugher$^{4}$,
J.~Garc\'ia-Bellido$^{29}$,
D.~W.~Gerdes$^{30,31}$,
D.~Gruen$^{18,19}$,
R.~A.~Gruendl$^{22,23}$,
J.~Gschwend$^{20,21}$,
W.~G.~Hartley$^{17,32}$,
D.~L.~Hollowood$^{33}$,
K.~Honscheid$^{34,35}$,
D.~J.~James$^{36}$,
K.~Kuehn$^{37}$,
M.~Lima$^{38,20}$,
M.~A.~G.~Maia$^{20,21}$,
R.~Miquel$^{39,24}$,
A.~A.~Plazas$^{40}$,
E.~Sanchez$^{26}$,
V.~Scarpine$^{4}$,
M.~Smith$^{41}$,
M.~Soares-Santos$^{42}$,
F.~Sobreira$^{43,20}$,
E.~Suchyta$^{44}$,
G.~Tarle$^{31}$,
A.~R.~Walker$^{13}$
	}
	\vspace{0.4cm}
	\\
	\parbox{\textwidth}{
		$^{1}$ Departamento de Ciencias Fisicas, Universidad Andres Bello Fernandez Concha 700, Las Condes, Santiago, Chile\\
		$^{2}$ Millennium Institute of Astrophysics, Chile\\
		$^{3}$ MIT Kavli Institute for Astrophysics and Space Research, Cambridge, MA 02139, USA\\
		$^{4}$ Fermi National Accelerator Laboratory, P. O. Box 500, Batavia, IL 60510, USA\\
		$^{5}$ Staples High School, Westport, CT 06880, USA\\
		$^{6}$ Institute of Astronomy, Madingley Road, Cambridge CB3 0HA, UK\\
		$^{7}$ CAPES Foundation, Ministry of Education of Brazil, Brasilia - DF 70040-020, Brazil\\
		$^{8}$ Department of Physics and Astronomy, PAB, 430 Portola Plaza, Box 951547, Los Angeles, CA 90095-1547, USA\\
		$^{9}$ European Southern Observatory, Karl-Schwarzschild-Strasse 2, 85748 Garching bei M\"unchen, DE\\
		$^{10}$ Department of Physics, University of California Davis, 1 Shields Avenue, Davis, CA 95616, USA\\
		$^{11}$ Instituto de F\'isica y Astronom\'ia, Universidad de Valpara\'iso, Avda. Gran Breta\~na 1111, Playa Ancha, Valpara\'iso 2360102, Chile\\
		$^{12}$ Subaru Telescope, National Astronomical Observatory of Japan, 650 N Aohoku Pl, Hilo, HI 96720, USA\\
		$^{13}$ Cerro Tololo Inter-American Observatory, National Optical Astronomy Observatory, Casilla 603, La Serena, Chile\\
		$^{14}$ Observatories of the Carnegie Institution of Washington, 813 Santa Barbara St., Pasadena, CA 91101, USA\\
		$^{15}$ CNRS, UMR 7095, Institut d'Astrophysique de Paris, F-75014, Paris, France\\
		$^{16}$ Sorbonne Universit\'es, UPMC Univ Paris 06, UMR 7095, Institut d'Astrophysique de Paris, F-75014, Paris, France\\
		$^{17}$ Department of Physics \& Astronomy, University College London, Gower Street, London, WC1E 6BT, UK\\
		$^{18}$ Kavli Institute for Particle Astrophysics \& Cosmology, P. O. Box 2450, Stanford University, Stanford, CA 94305, USA\\
		$^{19}$ SLAC National Accelerator Laboratory, Menlo Park, CA 94025, USA\\
		$^{20}$ Laborat\'orio Interinstitucional de e-Astronomia - LIneA, Rua Gal. Jos\'e Cristino 77, Rio de Janeiro, RJ - 20921-400, Brazil\\
		$^{21}$ Observat\'orio Nacional, Rua Gal. Jos\'e Cristino 77, Rio de Janeiro, RJ - 20921-400, Brazil\\
		$^{22}$ Department of Astronomy, University of Illinois at Urbana-Champaign, 1002 W. Green Street, Urbana, IL 61801, USA\\
		$^{23}$ National Center for Supercomputing Applications, 1205 West Clark St., Urbana, IL 61801, USA\\
		$^{24}$ Institut de F\'{\i}sica d'Altes Energies (IFAE), The Barcelona Institute of Science and Technology, Campus UAB, 08193 Bellaterra (Barcelona) Spain\\
		$^{25}$ Department of Physics and Astronomy, University of Pennsylvania, Philadelphia, PA 19104, USA\\
		$^{26}$ Centro de Investigaciones Energ\'eticas, Medioambientales y Tecnol\'ogicas (CIEMAT), Madrid, Spain\\
		$^{27}$ George P. and Cynthia Woods Mitchell Institute for Fundamental Physics and Astronomy, and Department of Physics and Astronomy, Texas A\&M University, College Station, TX 77843,  USA\\
		$^{28}$ Department of Physics, IIT Hyderabad, Kandi, Telangana 502285, India\\
		$^{29}$ Instituto de Fisica Teorica UAM/CSIC, Universidad Autonoma de Madrid, 28049 Madrid, Spain\\
		$^{30}$ Department of Astronomy, University of Michigan, Ann Arbor, MI 48109, USA\\
		$^{31}$ Department of Physics, University of Michigan, Ann Arbor, MI 48109, USA\\
		$^{32}$ Department of Physics, ETH Zurich, Wolfgang-Pauli-Strasse 16, CH-8093 Zurich, Switzerland\\
		$^{33}$ Santa Cruz Institute for Particle Physics, Santa Cruz, CA 95064, USA\\
		$^{34}$ Center for Cosmology and Astro-Particle Physics, The Ohio State University, Columbus, OH 43210, USA\\
		$^{35}$ Department of Physics, The Ohio State University, Columbus, OH 43210, USA\\
		$^{36}$ Harvard-Smithsonian Center for Astrophysics, Cambridge, MA 02138, USA\\
		$^{37}$ Australian Astronomical Observatory, North Ryde, NSW 2113, Australia\\
		$^{38}$ Departamento de F\'isica Matem\'atica, Instituto de F\'isica, Universidade de S\~ao Paulo, CP 66318, S\~ao Paulo, SP, 05314-970, Brazil\\
		$^{39}$ Instituci\'o Catalana de Recerca i Estudis Avan\c{c}ats, E-08010 Barcelona, Spain\\
		$^{40}$ Jet Propulsion Laboratory, California Institute of Technology, 4800 Oak Grove Dr., Pasadena, CA 91109, USA\\
		$^{41}$ School of Physics and Astronomy, University of Southampton,  Southampton, SO17 1BJ, UK\\
		$^{42}$ Brandeis University, Physics Department, 415 South Street, Waltham MA 02453\\
		$^{43}$ Instituto de F\'isica Gleb Wataghin, Universidade Estadual de Campinas, 13083-859, Campinas, SP, Brazil\\
		$^{44}$ Computer Science and Mathematics Division, Oak Ridge National Laboratory, Oak Ridge, TN 37831\\
	}
}

\date{Accepted XXX. Received YYY; in original form ZZZ}

\pubyear{2018}
\begin{document}
\label{firstpage}
\pagerange{\pageref{firstpage}--\pageref{lastpage}}
\maketitle

\begin{abstract}
We report upon the follow up of 34 candidate lensed quasars found in the Dark Energy Survey using NTT-EFOSC, Magellan-IMACS, KECK-ESI and SOAR-SAMI. These candidates were selected by a combination of double component fitting, morphological assessment and color analysis.  Most systems followed up are indeed composed of at least one quasar image and 13 with two or more quasar images: two lenses, four projected binaries and seven Nearly Identical Quasar Pairs (NIQs). The two systems confirmed as genuine gravitationally lensed quasars are one quadruple at $z_s=1.713$ and one double at $z_s=1.515$. Lens modeling of these two systems reveals that both systems require very little contribution from the environment to reproduce the image configuration. Nevertheless, small flux anomalies can be observed in one of the images of the quad. Further observations of 9 inconclusive systems (including 7 NIQs) will allow to confirm (or not) their gravitational lens nature.
\end{abstract}

\begin{keywords}
gravitational lensing: strong -- techniques: image processing -- surveys -- quasars: general
\end{keywords}

\section{Introduction}

The space and time distortion produced by a massive galaxy in close projection to the line of sight of a distant object, may produce multiple images of the background source. While this was predicted/theorized by \cite{einstein1936} and \cite{zwicky1937}, it took over four decades to discover the first lensed quasar \cite{walsh1979}.
Gravitationally lensed quasar systems are exceptional astrophysical and cosmological laboratories \cite[e.g.][]{CSS02}, and as such, have been sought after ever since.

Unfortunately, lensed quasars systems are very rare phenomena, since they require the chance alignment of a (rare) quasar with a (rare) foreground massive deflector. Their density on the sky is estimated to be a tenth per square degree or less, at the typical limit of present and upcoming surveys \citep{O+M10}. Thus finding lensed quasars requires wide area datasets and advanced techniques to sift through the large number of potential contaminants and false positives \citep[e.g.,][]{Ogu++06,Bro++03}. Up until now, of order a couple of hundred lensed quasars are known, including a couple of dozens of quadruply-imaged systems, which are the most valuable for many applications owing to their high information content.  Therefore, most applications of lensed quasars are limited by sample size.

The current generation of wide field imaging surveys provides an
opportunity to dramatically expand the samples of known lens quasars
and thus benefit all of their scientific applications. The STRong lensing
Insights into the Dark Energy Survey (STRIDES;
\url{strides.astro.ucla.edu}) collaboration \citep{agnello2015,treu2018} was formed to find
gravitationally lensed quasars in the Dark Energy Survey
\citep[][henceforth DES]{DES2016} with three broad ultimate goals:
analysis of the dark matter content of the lensing galaxies \citep[e.g.,][]{Sch++14}, analysis
of the structure of the lensed quasars \citep[e.g.,][]{Ang++08}, and measurement of a ``local''
Hubble constant via time delays \citep[e.g.,][]{Bon++17}.

Different groups within the STRIDES collaboration adopted complementary approaches
to identifying lensed quasar systems, In all but a few cases the DES
data alone do not suffice to classify a system as a ``secure'' lens
and follow-up observations are needed.  \cite{treu2018} give
an overview of the 2016 follow-up campaign and report on results from
two of the identification techniques.  Here we report results from a
third such technique.  \cite{ostrovski2018} reports results
from a fourth technique. 

While all four techniques begin with the some kind of color selection,
in the present approach that criterion is very broad, including many
hundreds of thousands of objects.  The second step is the automated
decomposition of the selected sources into pairs of objects, which are
then evaluated on morphological and refined color criteria to produce
candidate lensed systems.  Operationally, the method ultimately
involved the excision and analysis of more than a million $grizY$
cutouts to produce a list of objects for visual inspection and
possible follow-up.

In \S2 we discuss the selection process.  In \S3 we describe the
follow-up spectroscopic and imaging observations.  In \S4 we
discuss two systems that can securely be classified as lensed quasars
and twenty-three systems that cannot.  In \S5 we discuss nine systems for
which the observations are inconclusive, including seven Nearly Identical
Quasar pairs (NIQs).

For all sections, a flat cosmology with $\Omega_m = 0.3$ and $H_0=70 kms^{-1}Mpc^{-1}$ was assumed unless otherwise specified.

\section {Sample Selection}
\label{sec:sample}
To select our sample, we began with the WISE \citep{wright2010}
catalog and extracted 12\arcsec x12\arcsec DES $grizY$ cutouts for objects
with $m_{4.5\mu m} < 14.45$ and $m_{3.6\mu m} - m_{4.5\mu m} > 0.7$ on
the Vega system.  \cite{stern2012} have shown that color selection produces a
sample of quasars of relatively high completeness and high purity.
The low resolution of the WISE survey constitutes a virtue for our
purposes: it ensures that the photometry does not suffer from partial
resolution of the multiple lensed quasar images, which are typically
separated by less than $2\arcsec$. As such, close separation systems are identified as single objects and magnitude errors that arise from systems in close proximity, that are not deblended and treated separately, are avoided.

Rather than excise cutouts from the co-added DES images, we used the single best image in each filter as gauged by the effective exposure time $t_{eff}$ (see Table \ref{tab:desdata}). $t_{eff}$ is the ratio between the actual exposure time and the exposure time necessary to achieve the same signal to noise for point sources in nominal conditions \citep{morganson2018}. To test if the object was indeed a multi-component source, it was then split into two
components as described by \cite{schechter2017}.  Magnitudes were
obtained for the two components by forcing the same splitting on all
filters, taking the two components to have a common quasi-Gaussian
surface brightness profile but deriving it from the pair itself,
rather than using the local PSF as determined by the DES pipeline.  If
the quasi-Gaussian is substantially larger than the local PSF, the
split object is rejected as a pair of galaxies.

When the signal-to-noise permitted, a second test was applied, using
separate quasi-Gaussians for each component. If area inside half maximum  of the quasi-Gaussians was more than two pixels greater than that of the local PSF in each of three
filters,  the pair was rejected.  These two morphological rejection
criteria run the risk of rejecting systems in which the light for the
lensing galaxy makes a significant contribution.  It was also found to
fail for roughly 20\% of the of the known quasars in the analyzed footprints, with two or three images being
treated as a single extended image.  This shortcoming was subsequently
addressed by splitting objects into triplets when possible.

For the surviving pairs, a linear fit was obtained for the flux
ratios (expressed in magnitudes) as a function of $\log \lambda$.  As
described by \cite{schechter2017} they were given scores based on
the slope of the fit and the scatter about the fit (hereafter, ufom).  The highest
scores were given to systems with small scatter and a slight slope in the flux ratio,
favoring systems in which the fainter component is redder.  Systems
with scores less than 0.3 were not carried forward.

Combined magnitudes were computed by adding the fluxes of the two components using the calibration parameters
produced by the DES pipeline; these were embedded in the fits headers. The $griz$ color combinations from these total magnitudes were then
analyzed using a Gaussian mixture model similar to that used by
\cite{ostrovski2017}.  Each pair was assigned relative probabilities
$p(s),$ $p(g)$ and $p(q)$ of having star, galaxy and quasar colors,
respectively.  Systems with $p(q) < 0.5$ were not carried forward (with a few exceptions as shown in Table \ref{tab:Conc}).

The grizY cutouts for those systems with scores greater than 0.3 and
$p(q) > 0.5$ were examined visually, as were all systems with scores
greater than 0.5.  Pairs were culled for a variety of somewhat
subjective reasons. A final sample of 54 candidates resulted from the selection.

The DES cutouts were analyzed in two groups, drawn, respectively, from
the first and second year footprints of the DES \citep{flaugher2005, sanchez2016}. Note, however, that all imaging data used in this work resulted from the Y3A1 processing pipeline. The details of the DES imaging of the systems presented in this paper are shown in Table \ref{tab:desdata}.

\begin{landscape}

\begin{table}[]
	\caption{DES imaging data of the 34 lens candidate systems followed up. Exposure numbers, effective exposure time (Teff) and average FWHM are shown for each band. Naming of the systems reflects the coordinates of the brightest component. Note that the effective exposure time (Teff) is the ratio between the actual exposure time of the image (90 seconds in the g, r, i and z bands, and 45 seconds in the Y band) and the exposure time necessary to achieve the same signal to noise for point sources in nominal conditions.}
	\label{tab:desdata} 
	\begin{center}
		\begin{tabular}{l ccccc ccccc ccccc}
			\hline
			ID                         &    \multicolumn{5}{c}{exposure number}     &       \multicolumn{5}{c}{Teff}        &                \multicolumn{5}{c}{FWHM}                   \\ \hline
			                         & g      &   r    &   i    &   z    &   Y    & g     &   r   &   i   &   z   &   Y   & g    &  r   &  i   &  z   &              Y                \\ \hline
			DESJ004306.87-411032.6 		& 241210 & 369439 & 372931 & 253452 & 384081 & 0.63 & 0.65 & 0.50   & 0.94  & 1.24  & 1.19 & 1.22 & 1.31 & 0.74 & 0.84 \\
			DESJ005301.91+002042.9     & 257490 & 377405 & * & 377746 & 238105 & 0.79  & 0.77  & *  & 0.89  & 1.07  & 1.08 & 1.16 & * & 0.87 &            0.97               \\
			DESJ005817.07-612004.6     & 387027 & 499372 & 370289 & 382172 & 470190 & 0.98  & 1.22  & 0.93  & 0.86  & 1.02  & 0.97 & 0.97 & 0.92 & 0.65 &            0.86               \\
			DESJ012006.38-435440.8     & 242801 & 242802 & 242803 & 256707 & 256708 & 1.05  & 0.92  & 1.03  & 1.03  & 1.26  & 1.32 & 1.28 & 1.11 & 0.92 &            0.90               \\
			DESJ013733.00-022242.6     & 240771 & 490665 & 372593 & 488823 & 384051 & 0.70  & 0.88  & 1.00  & 1.06  & 1.45  & 1.17 & 0.98 & 0.82 & 0.92 &            0.76               \\
			DESJ021524.22-472845.2     & 261270 & 258908 & 257598 & 257597 & 253506 & 0.92  & 1.22  & 0.97  & 1.05  & 1.36  & 0.92 & 0.92 & 0.84 & 0.93 &            0.79               \\
			DESJ023004.58-070445.9     & 363888 & 363890 & 363891 & 151376 & 349419 & 1.06  & 1.13  & 1.56  & 0.95  & 1.23  & 0.94 & 1.16 & 1.10 & 0.95 &            0.97               \\
			DESJ024018.40-020850.2     & 232436 & 361677 & 363897 & 253437 & 253418 & 0.87  & 0.87  & 1.20  & 0.96  & 1.15  & 1.07 & 1.02 & 1.06 & 0.79 &            0.71               \\
			DESJ025357.76-050454.7     & 277299 & 361681 & 361682 & 253449 & 253450 & 0.69  & 1.08  & 1.78  & 1.00  & 1.28  & 1.11 & 1.14 & 1.06 & 0.91 &            0.76               \\
			DESJ032559.42-451820.9     & 269629 & 269980 & 268015 & 513378 & 513379 & 1.00  & 0.86  & 1.08  & 1.10  & 1.42  & 1.03 & 1.09 & 0.98 & 0.77 &            0.77               \\
			DESJ040235.79-152328.4     & 363934 & 363938 & 363939 & 360692 & 359745 & 1.07  & 1.16  & 1.50  & 1.32  & 1.25  & 0.97 & 1.13 & 1.08 & 1.04 &            0.74               \\
			DESJ040559.80-330851.4     & 511795 & 511794 & 511793 & 403055 & 403823 & 0.98  & 1.27  & 1.65  & 2.02  & 3.05  & 0.95 & 1.09 & 0.99 & 1.10 &            0.67               \\
			DESJ040710.22-500600.9     & 268872 & 268873 & 257650 & 257651 & 266988 & 1.17  & 1.34  & 1.19  & 1.78  & 1.33  & 0.97 & 1.10 & 0.79 & 0.78 &            0.75               \\
			DESJ042553.23-453935.3     & 167349 & 167350 & 269672 & 263618 & 266247 & 1.11  & 1.15  & 1.24  & 1.00  & 1.12  & 1.03 & 1.02 & 0.92 & 0.89 &            0.78               \\
			DESJ043857.14-341609.9     & 402697 & 402339 & 403095 & 395561 & 386819 & 0.73  & 0.88  & 2.03  & 1.02  & 1.75  & 1.32 & 1.14 & 0.91 & 0.99 &            0.89               \\
			DESJ044042.84-200818.8     & 502762 & 502761 & 363955 & 403845 & 497417 & 1.07  & 1.18  & 1.43  & 1.15  & 1.58  & 1.00 & 0.96 & 0.98 & 0.84 &            0.81               \\
			DESJ044316.64-330845.1     & 500877 & 403096 & 403098 & 386431 & 386432 & 0.99  & 1.44  & 1.66  & 1.07  & 1.50  & 1.09 & 0.89 & 0.92 & 0.83 &            0.78               \\
			DESJ044402.20-371831.7     & 401196 & 381959 & 401938 & 386434 & 283819 & 1.33  & 0.92  & 1.67  & 1.05  & 0.91  & 0.93 & 1.08 & 0.82 & 0.87 &            0.77               \\
			DESJ045137.23-341006.0     & 500537 & 403101 & 402341 & 386820 & 395564 & 1.68  & 1.27  & 1.32  & 1.53  & 1.45  & 0.73 & 0.95 & 1.16 & 0.87 &            0.89               \\
			DESJ050912.77-235049.3     & 403481 & 500531 & 407990 & 404804 & 403873 & 0.89  & 2.84  & 1.34  & 0.85  & 1.47  & 1.05 & 0.82 & 0.83 & 0.92 &            0.77               \\
			DESJ051656.90-602031.4     & 270361 & 270364 & 270365 & 181024 & 405234 & 0.97  & 1.03  & 1.13  & 0.85  & 1.51  & 1.06 & 0.92 & 0.89 & 0.87 &            0.67               \\
			DESJ054430.63-592238.7     & 269310 & 515751 & 179958 & 267039 & 267040 & 0.73  & 0.88  & 0.52  & 0.78  & 1.12  & 1.12 & 0.98 & 1.02 & 0.84 &            0.82               \\
			DESJ060003.77-284758.5     & 271655 & 389583 & 390625 & 497865 & 408766 & 1.12  & 1.18  & 1.12  & 1.02  & 0.97  & 0.90 & 1.00 & 0.99 & 0.76 &            0.89               \\
			DESJ204726.90-480147.9     & 360503 & 367449 & 239553 & 487035 & 244132 & 0.65  & 0.61  & 0.80  & 0.53  & 1.06  & 1.25 & 1.21 & 0.90 & 0.97 &            0.82               \\
			DESJ210358.13-580049.0     & 230032 & 231072 & 231073 & 242675 & 479334 & 0.88  & 1.30  & 1.49  & 1.31  & 0.90  & 1.20 & 0.81 & 0.78 & 0.74 &            1.04               \\
			DESJ214148.85-462945.7     & 367473 & 371363 & 362359 & 364190 & 364625 & 0.65  & 0.91  & 2.07  & 0.81  & 1.13  & 1.24 & 0.86 & 0.74 & 0.82 &            0.93               \\
			DESJ215426.50-441044.4     & 371369 & 362366 & 474260 & 372006 & 372060 & 0.94  & 1.52  & 1.44  & 2.11  & 1.50  & 0.83 & 0.97 & 0.90 & 0.66 &            0.88               \\
			DESJ220039.00-471900.0     & 370205 & 367478 & 482106 & 372053 & 244206 & 0.62  & 0.68  & 1.10  & 0.95  & 0.96  & 1.11 & 1.14 & 0.85 & 0.95 &            0.94               \\
			DESJ221710.62+013808.3     & 242443 & 242429 & 233497 & 233499 & 479364 & 0.56  & 0.85  & 0.81  & 1.05  & 0.69  & 1.16 & 0.93 & 1.04 & 1.01 &            1.21               \\
			DESJ225007.92-604723.1     & 350159 & 362394 & 370229 & 374815 & 374808 & 1.36  & 1.00  & 0.80  & 0.95  & 1.20  & 0.87 & 0.95 & 0.94 & 0.72 &            0.70               \\
			DESJ230329.90-484430.5     & 233587 & 362393 & 362392 & 243157 & 355343 & 0.95  & 1.82  & 2.62  & 1.29  & 1.34  & 1.15 & 0.93 & 0.94 & 0.86 &            0.92               \\
			DESJ230602.30-565755.5     & 231544 & 350166 & 239693 & 466778 & 348401 & 0.84  & 1.41  & 1.60  & 1.07  & 2.67  & 1.27 & 0.90 & 0.96 & 0.88 &            0.74               \\
			DESJ232625.74-480548.5     & 350889 & 267575 & 475865 & 470026 & 469536 & 1.12  & 0.90  & 1.18  & 2.05  & 1.94  & 1.19 & 1.10 & 1.13 & 0.90 &            1.00               \\
			DESJ233713.66+005610.8     & 232371 & 239622 & 239639 & 242741 & 238084 & 0.59  & 0.95  & 1.27  & 2.35  & 1.02  & 1.24 & 1.01 & 0.97 & 0.78 &            1.09             \\
		\end{tabular}	                                                                   
		
	\end{center}    
    $*$No i band available.\\                                                                                                                                                         
\end{table}                                                                               

\end{landscape}

\section{Follow Up}
\label{sec:follup}

Thirty-four of the selected systems were followed up. Out of them, thirty-two spectroscopically. Twenty-two of them with NTT EFOSC2, 8 with ESI and 2 with IMACS. Table \ref{tab:Conc} indicates the data obtained for each followed up system. The details of each observing run are given in the 2016 STRIDES campaign overview paper \citep{treu2018}, however, for completeness we describe them here as well. Table \ref{tab:Conc} shows the selection and follow up parameters of all systems. 

Note that for all spectroscopic observations, the slit angle was not defined at the parallactic angle but by the angle defined between the multiple lensed point-like images. As such, significant slit losses are expected in the spectra. Nevertheless, the validity of the flux ratio between multiple images should hold. Due to the nature of the observations, extraction was performed with custom routines: two Gaussian profiles were fitted wavelength wise on the data iteratively until residuals were minimized.

In addition, higher resolution imaging (than that provided by DES) was obtained for 11 systems (two of them without spectroscopic follow up). Eight were observed with the 4.1m Southern Astrophysical Research Telescope (SOAR) SAMI instrument at Cerro Pachon with its Adaptive Optics (AO) system SAM \citep{tokovinin2016}. Imaging was carried out in the redder SAMI bands to maximize AO correction and optimize the contrast between quasar and deflector galaxy. The pixel scale was 0.09\arcsec/pix (2$\times$2 binning of 0.045\arcsec/pix pixels) and the typical exposure time was 3x180 seconds. Four systems were observed with Magellan IMACS in its imaging mode.

\begin{table*}[tb]
	\caption{The 34 lensed candidate systems followed up, split into ``Conclusive'' and ``Inconclusive'' (those whose nature has been confirmed and those that it has not) as well as their classification. The columns show respectively: Name (ID), g band magnitude of the faintest image (g), separation between the two images of the system in arcseconds ($\Delta \theta$), magnitude difference between the faintest and brightest image, respectively (g$_{\rm dif}$), ufom score as described in the text (ufom), quasar color probability as described in the text ($p(q)$), spectroscopic follow instrument (Spec.), imaging follow-up instrument (Ima.) and measured quasar redshift if available (redshift). Redshift uncertainties are $\sigma_z \lesssim 0.001$}
	\label{tab:Conc} 
	\begin{center}
		\begin{tabular}{lllllllllllll}
			\hline
			&			                                                      & ID             & g     & $\Delta \theta$[\arcsec] & g$_{\rm dif}$ & ufom      & $p(q)$    & Spec.  & Ima.        & redshift                         \\ \hline
			\multirow{25}{*}{\rotatebox[origin=c]{90}{Conclusive}} &			\multirow{2}{*}{\rotatebox[origin=c]{90}{Lens}}       & DESJ 0405-3308 & 22.25 & -                        & -0.84         & 1.5       & 4e-3$^a$  & IMACS  & SAMI, IMACS & z=1.713                          \\
			&			                                                      & DESJ 0407-5006 & 19.63 & 1.69                     & -1.27         & 0.67      & 0.76      & EFOSC2 & IMACS       & z=1.515                          \\ \cline{2-11}
			&			\multirow{4}{*}{\rotatebox[origin=c]{90}{Proj. Bin.}} & DESJ 0215-4728 & 19.99 & 1.07                     & -1.16         & 0.43      & 0.60      & EFOSC2 &             & z$\rm _1$=1.692, z$\rm _2$=0.467 \\
			&			                                                      & DESJ 0240-0208 & 19.22 & 0.94                     & -0.48         & 0.3       & 0.72      & ESI    &             & z$\rm _1$=1.685, z$\rm _2$=1.058 \\
			&			                                                      & DESJ 0425-4539 & 20.73 & 1.90                     & -1.08         & 0.41      & 2e-6$^b$  & EFOSC2 &             & z$\rm _1$=1.020, z$\rm _2$=0.913 \\
			&			                                                      & DESJ 2303-4844 & 20.93 & 2.21                     & -1.52         & 0.34      & 0.95      & EFOSC2 &             & z$\rm _1$=2.020, z$\rm _2$=1.164 \\  \cline{2-11}
			&			\multirow{7}{*}{\rotatebox[origin=c]{90}{QSO+ Star}}  & DESJ 0053+0020 & 18.78 & 1.12                     & -0.85         & 0.62      & -----$^c$ & ESI    &             & z=1.320                          \\
			&			                                                      & DESJ 0325-4518 & 21.14 & 2.13                     & -1.44         & 0.33      & 0.60      & EFOSC2 &             & z=0.695                          \\
			&			                                                      & DESJ 0516-6020 & 19.65 & 2.52                     & -1.21         & 0.67      & 0.97      & EFOSC2 &             & z=1.039                          \\
			&			                                                      & DESJ 2103-5800 & 20.08 & 2.15                     & -0.67         & 0.37      & 0.02$^d$  & EFOSC2 &             & z=0.905                          \\
			&			                                                      & DESJ 2154-4410 & 19.50 & 1.94                     & -1.84         & 0.51      & 0.73      & EFOSC2 &             & z=1.750                          \\
			&			                                                      & DESJ 2217+0138 & 20.93 & 1.84                     & -1.82         & 0.38      & 0.95      & EFOSC2 &             & z=1.693                          \\
			&			                                                      & DESJ 2306-5657 & 19.92 & 2.17                     & -1.74         & 0.35      & 0.87      & EFOSC2 &             & z=1.010                          \\  \cline{2-11}
			&			\multirow{4}{*}{\rotatebox[origin=c]{90}{QSO+Gal.}}   & DESJ 0043-4110 & 21.91 & 3.17                     & -1.81         & 0.53      & 0.91      & EFOSC2 &             & z=0.882                          \\
			&			                                                      & DESJ 0438-3416 & 20.77 & 2.09                     & -1.60         & 0.46      & 0.92      & EFOSC2 &             & z=1.097                          \\
			&			                                                      & DESJ 0444-3718 & 20.50 & 3.53                     & -2.08         & 0.39      & 0.77      & EFOSC2 &             & z=1.345                          \\
			&			                                                      & DESJ 2326-4805 & 20.94 & 1.84                     & -2.36         & 0.59      & 0.70      & EFOSC2 &             & z=1.364                          \\  \cline{2-11}
			&			\multirow{8}{*}{\rotatebox[origin=c]{90}{Em. Gal.}}   & DESJ 0137-0222 & 20.31 & 0.85                     & -0.27         & 0.39      & 0.52      &        & SAMI        &  \\
			&			                                                      & DESJ 0253-0504 & 20.59 & 0.95                     & -1.32         & 0.46      & 0.58      & ESI    &             &  \\
			&			                                                      & DESJ 0402-1523 & 20.23 & 1.87                     & -0.60         & 0.36      & 0.79      & ESI    &             &  \\
			&			                                                      & DESJ 0440-2008 & 19.23 & 1.49                     & -0.97         & 0.33      & 0.56      & ESI    &             &  \\
			&			                                                      & DESJ 0443-3308 & 19.58 & 0.91                     & -2.84         & 0.30      & 0.96      &        & SAMI        &  \\
			&			                                                      & DESJ 0451-3410 & 21.39 & 0.99                     & -1.90         & 0.34      & 0.94      & EFOSC2 &             &  \\
			&			                                                      & DESJ 0600-2847 & 20.97 & 0.98                     & -1.47         & 0.31      & 0.81      & ESI    &             &  \\
			&			                                                      & DESJ 2047-4801 & 20.14 & 3.70                     & -0.87         & 0.56      & 0.91      & IMACS  &             &  \\ \hline
			\multirow{9}{*}{\rotatebox[origin=c]{90}{Inconclusive}}&			\multirow{7}{*}{\rotatebox[origin=c]{90}{NIQ}}        & DESJ 0058-6120 & 20.10 & 3.03                     & -2.49         & 0.43      & 0.95      & EFOSC2 & SAMI        & z=1.322                          \\
			&			                                                      & DESJ 0120-4354 & 19.97 & 0.84                     & -0.15         & 0.38      & 0.86      & EFOSC2 &             & z=1.910                          \\
			&			                                                      & DESJ 0544-5922 & 18.93 & 1.24                     & -0.15         & 0.41      & 0.93      & EFOSC2 & IMACS       & z=1.319                          \\
			&			                                                      & DESJ 2141-4629 & 20.46 & 0.91                     & -0.54         & -----$^e$ & 0.55      & IMACS  & SAMI        & z=1.762                          \\
			&			                                                      & DESJ 2200-4719 & 19.78 & 3.60                     & -2.09         & 0.53      & 0.94      & EFOSC2 &             & z=1.608                          \\
			&			                                                      & DESJ 2250-6047 & 21.02 & 2.02                     & -1.31         & 0.46      & 0.70      & EFOSC2 & SAMI        & z=1.080                          \\
			&			                                                      & DESJ 2337+0056 & 20.39 & 1.35                     & -0.91         & 0.79      & 0.93      & EFOSC2 & IMACS       & z=0.710                          \\ \cline{2-11}
			&			\multirow{2}{*}{\rotatebox[origin=c]{90}{Other}}      & DESJ 0230-0704 & 18.05 & 0.57                     & -0.70         & 0.37      & 0.67      & ESI    & SAMI        & QSO z=2.01 + point source        \\
			&			                                                      & DESJ 0509-2350 & 19.94 & 1.99                     & -2.16         & 0.38      & 0.75      & ESI    & SAMI        & QSO z=2.08 + point source        \\ \hline
		\end{tabular}	                                                                   
		
	\end{center}                                                                           
	$^a$Flagged for observation due to quad nature, despite low p(q).\\                      
	$^b$Flagged for observation due to the strong photometric variability observed, despite low p(q).\\
	$^c$No i band available, precluding a p(q) measurement.\\                                
	$^d$Flagged for observation due to fairly blue color (g-i=0.18), despite low p(q).\\
	$^e$g band fitting failed due to large seeing so no ufom measurement available. g magnitude difference extrapolated from the r, i and z magnitude differences.\\ 	                                                                                       
\end{table*}

\section{Conclusive Systems}
\label{sec:confirmed}
                       
Twenty-five systems have had their nature confirmed by our spectroscopic follow up. A summary of their properties and follow up data is shown in Table \ref{tab:Conc}. 

\subsection{Lenses}

Two systems have been confirmed as genuine gravitationally lensed quasars. These two systems, besides having identical spectra of the two components, show evidence of a galaxy between them after subtraction of the quasar images. However no absorption lines are identified in the spectra of the systems, so no redshift measurement is available for their lensing galaxies. For these two systems we have performed PSF fitting to obtain the astrometry of the quasar images and lens galaxy in order to construct lens models.  The PSF fitting was carried out using a purpose-built program that incorporates subroutines from the program DoPHOT \citep{schechter1993}.  It creates a tabulated PSF using a nearby star and simultaneously fits the tabulated PSF to the quasar images and a quasi-Gaussian to the central object. The lens models were performed using \textsc{glafic} \citep{oguri2010} and isothermal mass profiles.

\subsubsection{DESJ 0405-3308}

\begin{figure}
	\includegraphics[width=\columnwidth]{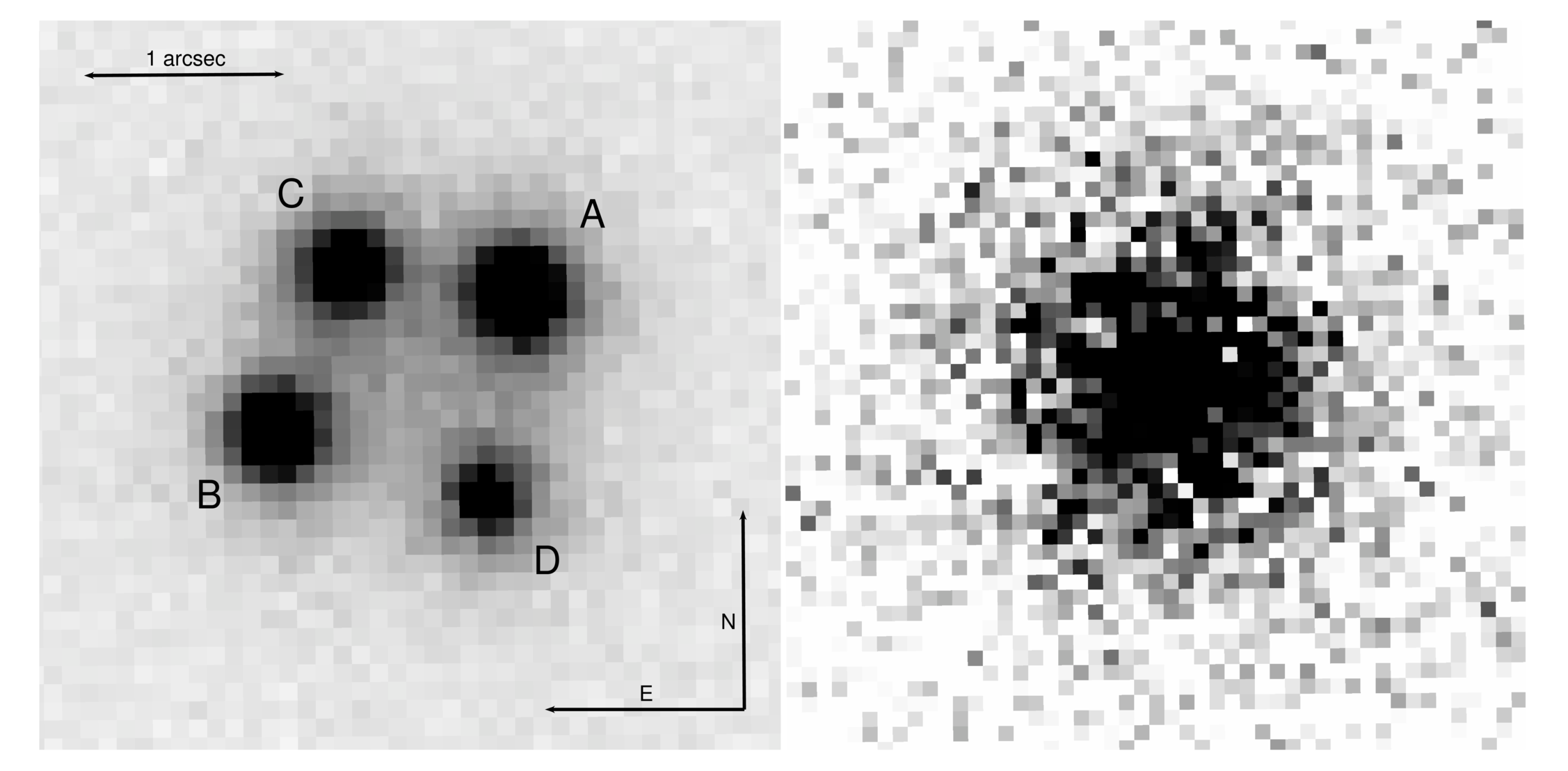}
	\caption{Left: DESJ 0405-3308 SOAR SAMI z image with naming convention. Right: contrast enhanced (10$\times$) version of the same image with PSFs of the four quasar images subtracted.}
	\label{0405im}
\end{figure}

\begin{figure}
	\includegraphics[width=\columnwidth]{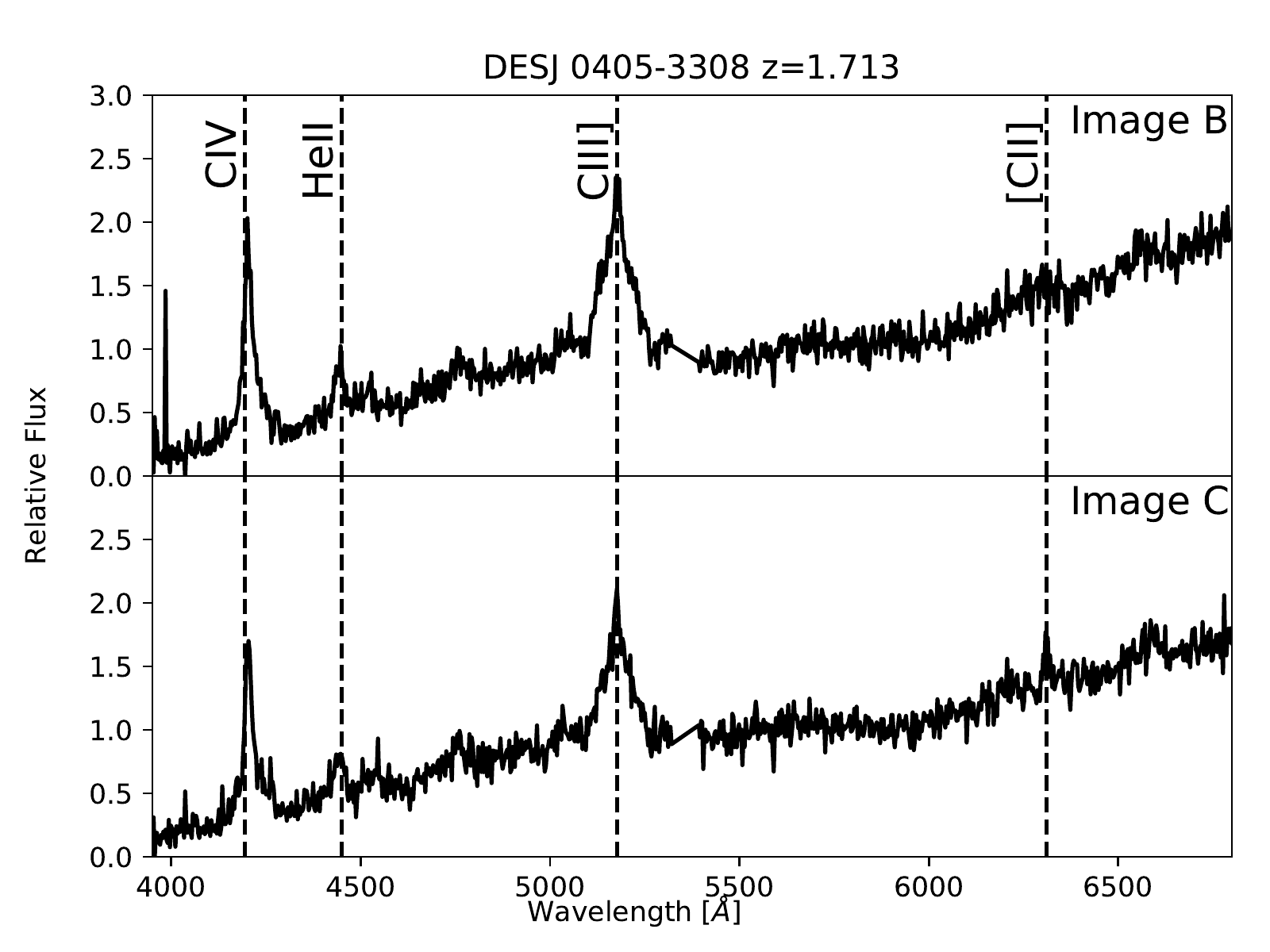}
    \caption{DESJ 0405-3308 Magellan IMACS spectra. Segmented lines show the identified emission lines used to measure the redshift shown above the figure panel.}
    \label{0405spec}
\end{figure}

DESJ 0405-3308 follow up imaging was obtained with Magellan IMACS in bands g, r and i on 2016 November 29 and with SOAR SAMI in bands R,V and z on 2016 December 3. The z band image was used for PSF fitting astrometry measurements. The photometry for the lensed images and galaxy (see Fig. \ref{0405im} for naming scheme) is shown in Table \ref{0405phot}. A single slit was used on Magellan IMACS through images B and C. Fig. \ref{0405spec} shows the extracted spectra and the identified lines at a redshift of z=1.713$\pm$0.001. The relative astrometry of the system is shown in Table \ref{0405ast}.

\begin{table}
	\caption{PSF fitting IMACS photometry (g, r and i) SAMI (z) for the DES J0405-3308. The position of lensed quasar images was fixed to positions fitted with the high resolution SAMI z image. Typical photometric uncertainty is 0.03 [mag]} 
	\begin{center}
		\begin{tabular}{ccccc}
			\hline
			Image &   g   &   r   &   i   &   z   \\ \hline
			  A   & 21.41 & 20.52 & 19.75 & 19.51 \\
			  B   & 21.54 & 20.60 & 19.87 & 19.62 \\
			  C   & 21.63 & 20.76 & 19.99 & 19.74 \\
			  D   & 22.25 & 21.15 & 20.36 & 20.00 \\
			  G   & 22.73 & 21.95 & 20.76 & 19.43 \\ \hline
		\end{tabular}
	\end{center}
	\label{0405phot}
\end{table}

\begin{table*}
	\caption{Observed and best fit modeled parameters for DESJ 0405-3308. Astrometry, flux ratios and time delays with respect to image B. Since we do not have the redshift of the lensing galaxy, time delays are scaled  such that $\Delta_t=\Delta_t^\star (1+z_l) \frac{D_L}{D_{LS}}$ days} 
	\label{0405ast}
	\centering
		\begin{tabular}{c|c|c|c|c|c|c|c|c|c|c|c|c|c|c|c|c|c|c|c}
			\hline
			Comp &    \multicolumn{3}{c}{Astrometry}   &  & \multicolumn{4}{c}{Flux Ratios} & &\multicolumn{7}{c}{Model output parameters} \\ \hline
			          & $\Delta$ RA & $\Delta$ DEC & error &  &  g   &  r   &  i   &  z   &  & $\Delta$ RA & $\Delta$ DEC & $\kappa$ &      $\gamma$  & $\mu$    &FR&  \multicolumn{1}{c}{$\Delta t^\star$} \\ \hline
			    A     &   -1.211    &    0.674     & 0.040 &  & 1.13 & 1.08 & 1.12 & 1.11 &  &   -1.209    &    0.669     &   0.45   &        0.45         & 9.30 & 1.06   &                   0.15                 \\
			    B     &     0.0     &     0.0      &  0.0  &  & 1.0  & 1.0  & 1.0  & 1.0  &  &     0.0     &     0.0      &   0.44   &        0.44         & 8.82 &   1.0   &                   0.0                  \\
			    C     &   -0.349    &    0.819     & 0.050 &  & 0.92 & 0.86 & 0.90 & 0.90 &  &   -0.363    &    0.822     &   0.55   &        0.55         &  -9.27 &  1.05   &                   1.02                 \\
			    D     &   -1.060    &    -0.311    & 0.070 &  & 0.52 & 0.60 & 0.64 & 0.71 &  &   -1.010    &    -0.343    &   0.58   &        0.58         & -6.05   &  0.69   &                   2.12                 \\
			 G   &   -0.680    &    0.216     & 0.120 &  &      &      &      &      &  &   -0.679    &    0.176     &          &                     & &   \\ \hline
		\end{tabular}
\end{table*}

In making the lens model, we have selected a Singular Isothermal Ellipsoid (SIE) for the mass profile and have obtained a good fit within the astrometric uncertainties. The constraints used in the lens model are only the relative position of the quasar images and the lens galaxy. The best fit SIE parameters obtained were an Einstein radius $R_E$=0\farcs69 and an ellipticity $e$=0.17 at a position angle $e_{PA}$=29$^\circ$. The secondary parameters obtained through the mass model are shown in Table \ref{0405ast}. The small ellipticity result for the mass profile fit is consistent with the almost circular shape of the lensing galaxy. As such, no additional external perturbation (shear) was required for the model. We do note, however, that the range of allowed ellipticies for our lens model is $0.15<e<0.55$ at a 90\% confidence level. Some rough consistency can be seen in the measured flux ratios (uncertainty smaller than 10\%) and the modeled flux ratios of images A and B, however, this is not the case with images C and D.  In the case of image C, the almost negligible wavelength dependence of the observed flux ratios suggest that difference between the observed and modeled ratios might be due to substructure in the lens or environmental contributions, especially considering that no emission line residual is observed in the spectroscopic flux ratio between images C and B. In image D, the observed flux ratio increases by a factor 1.4 from the g to the z band. This chromatic variation could be explained by  as well as differential dust extinction projected in front of image D \citep[see e.g.][]{anguita2008b,yonehara2008}. We note that both image C and D are saddle point images in the lens model and thus more prone to microlensing flux fluctuations \citep{schechter2002}. Exploring these anomalies goes beyond the scope of this paper. Further space based and/or adaptive optics imaging along deeper spectroscopy will allow more detailed mass and flux models for the system.

\begin{figure}
	\includegraphics[width=\columnwidth]{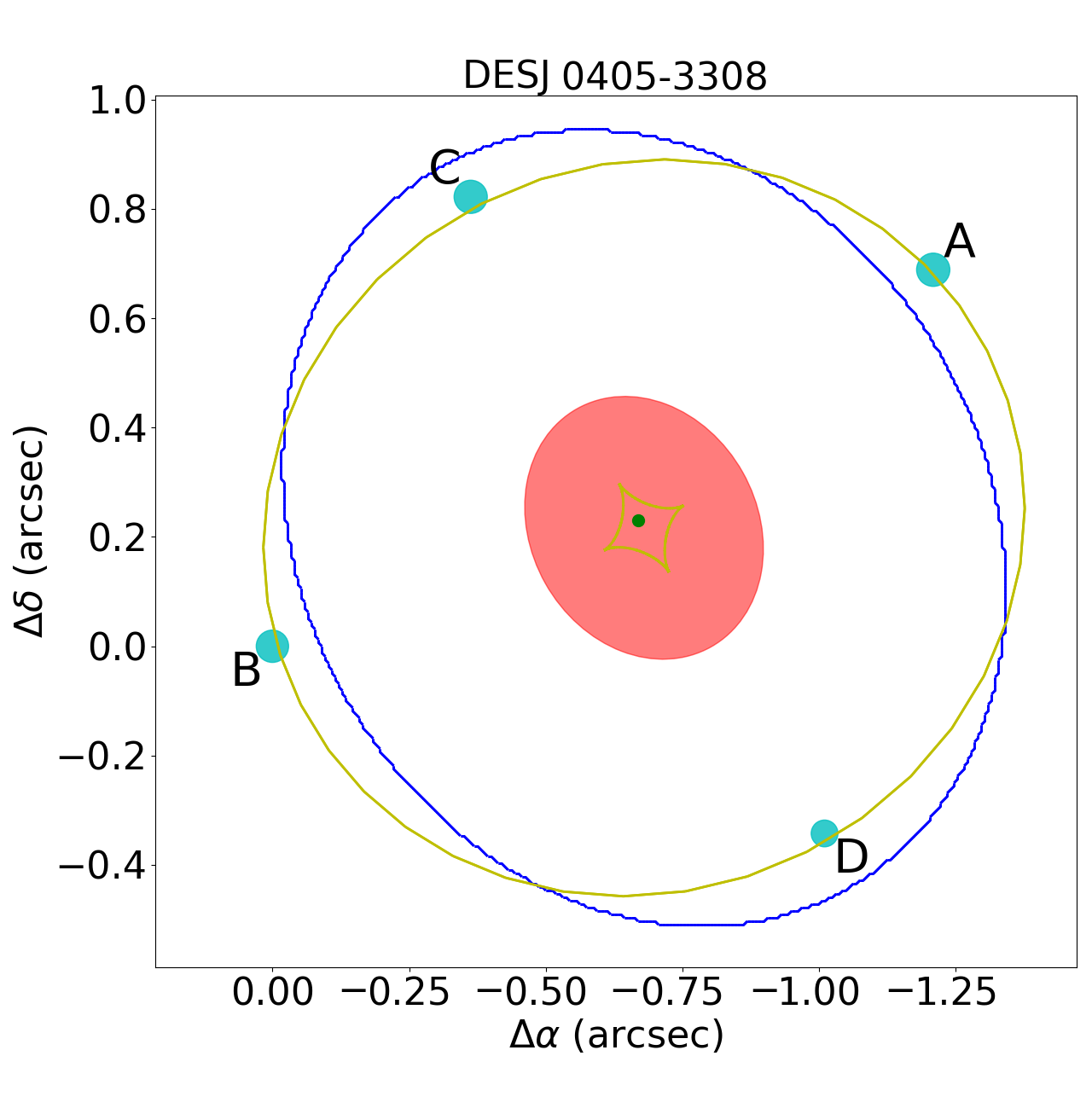}
	\caption{DESJ 0405-3308 lens model. Caustics are shown in yellow and the critical curve in blue. The source quasar is shown in green and the lens galaxy in red.}
	\label{0405mod}
\end{figure}

\subsubsection{DESJ 0407-5006}

\begin{figure}
	\includegraphics[width=\columnwidth]{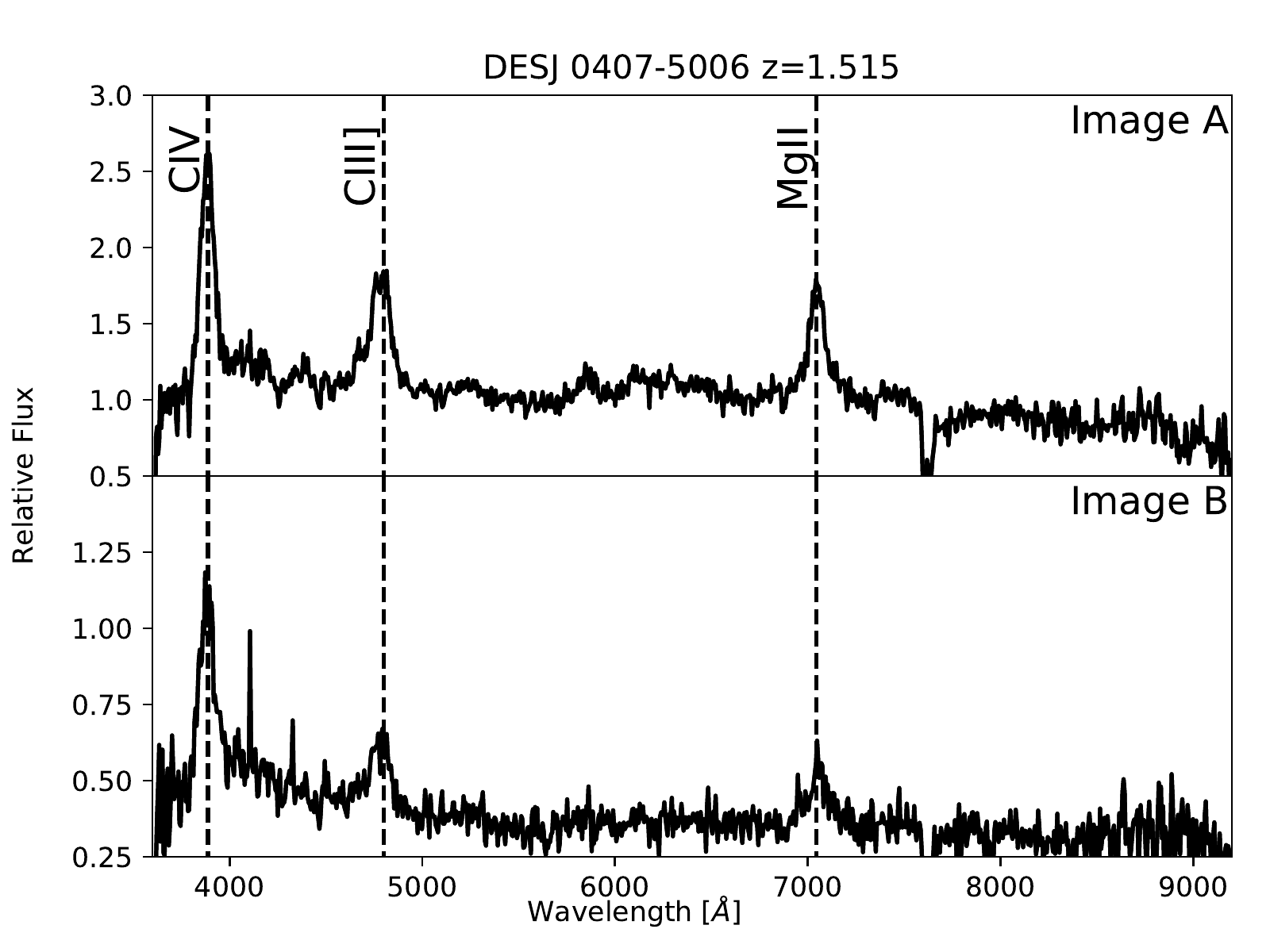}
    \caption{DESJ 0407-5006 NTT EFOSC2 spectra.  Segmented lines show the identified emission lines used to measure the redshift shown above the figure panel.}
    \label{0407spec}
\end{figure}

Follow up imaging for DESJ 0407-5006 was obtained with Magellan IMACS in the i-band on 2016 December 1. The photometry for the lensed images and galaxy is shown in Table \ref{0407phot}. Follow-up spectra were obtained with NTT-EFOSC2, as described above, with an exposure time of 600s. Fig. \ref{0407spec} shows the extracted spectra and the identified emission lines at a redshift of z=1.515$\pm$0.001. A slight emission line residual and chromaticity can be observed in the spectroscopic flux ratio. Although the significance of the signal does not allow us to draw any conclusions regarding microlensing in this system, it cannot be ruled out. Using the IMACS data and our PSF fitting technique we obtain the relative astrometry shown in Table \ref{0407ast}.

\begin{table}
	\caption{Photometry for the DES J0407-5006. No galaxy was detected in the DES imaging so no photometry for the lensing galaxy is available. As the lensing galaxy was not fitted in the DES imaging, the flux of the lensed images are subject to contamination by it. Typical photometric uncertainty for Magellan and DES imaging are respectively 0.03 and 0.06 [mag].} 
	\begin{center}
		\begin{tabular}{cccccc}
			\hline
			Image & \multicolumn{4}{c}{DES} & IMACS   \\ \hline
			      &          g          &   r   &   i   &   z   & i     \\ \hline
			  A   &        18.35        & 18.07 & 18.00 & 18.10 & 18.01 \\
			  B   &        19.62        & 19.36 & 19.19 & 19.19 & 19.26 \\
			  G   &          -          &   -   &   -   &   -   & 19.97 \\ \hline
		\end{tabular}
	\end{center}
	\label{0407phot}
\end{table}

The light profile fit shows a nearly circular shape for the lensing galaxy. Given the reduced number of positional constraints of this double system, we have selected a singular isothermal sphere plus external shear as a mass model and we have added the spectroscopic flux ratio measured on top of the CIII] and MgII emission lines as a constraint (we avoid using the CIV since it is very close to the edge of the CCD) to the relative positions of the quasar images and the lens galaxy. We conservatively measure a flux ratio of A/B=3.0$\pm$0.2, which is consistent with the observed i band flux ratio. The best fit requires Einstein radius of $R_E$=0\farcs87, together with a small external shear ($\gamma<0.06$ at 90\% confidence) at 130 degrees measured north to east (or south to west). The secondary parameters obtained through the mass model are shown in Table \ref{0407ast}. We note that due to the very small shear, even when the best fit position angle is at 130 degree east from north, the allowed range for this direction is poorly constrained to $\pm$45 degrees. As with DESJ 0405-3308, further space based and/or adaptive optics imaging along with deeper spectroscopy will allow more detailed mass and flux models of the system.

\begin{figure}
	\includegraphics[width=\columnwidth]{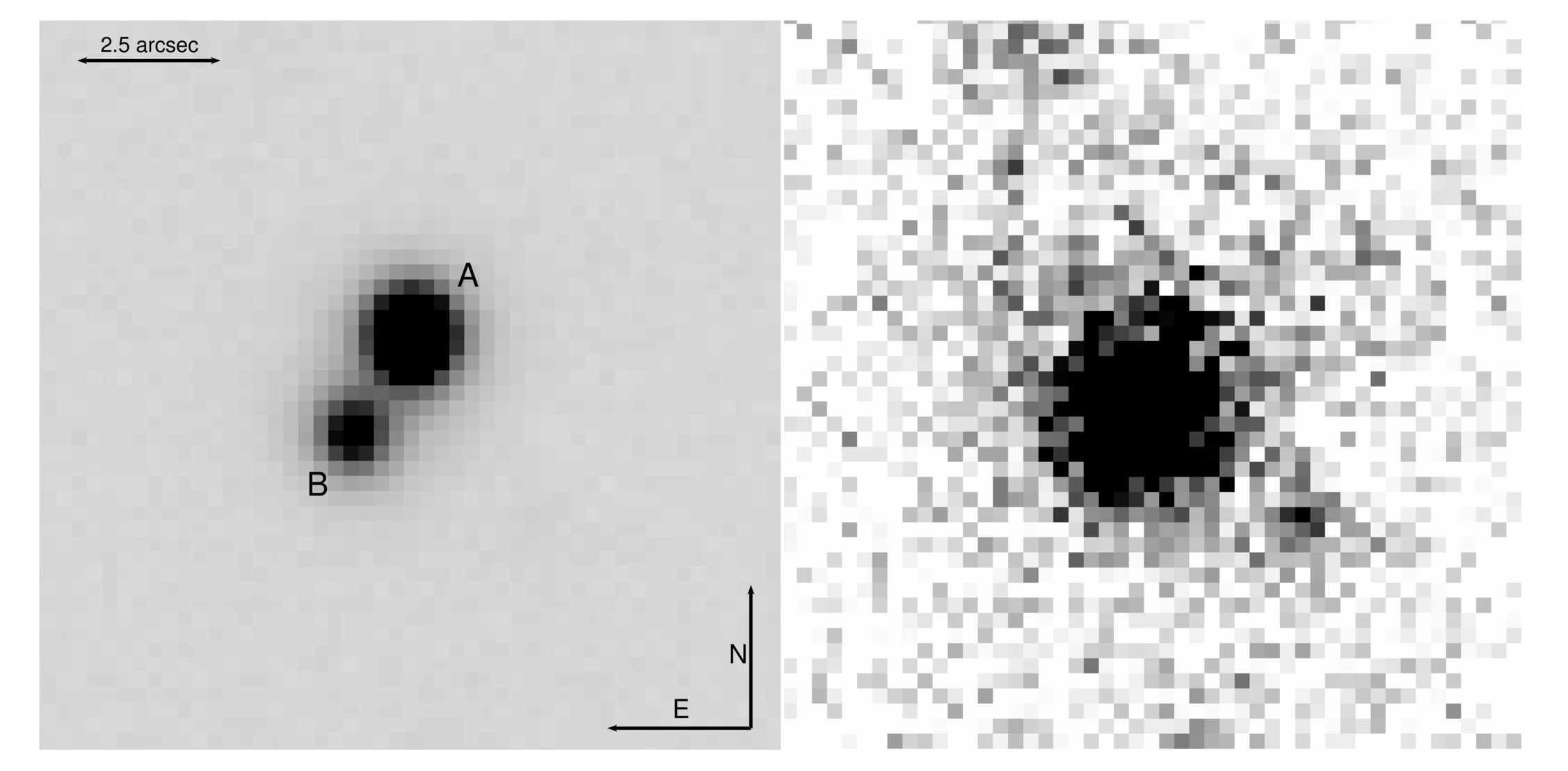}
	\caption{Left: DESJ 0407-5006 Magellan IMACS i image with the naming convention. Right: contrast enhanced (30$\times$) version of the same image with PSFs of the two quasar images subtracted.}
	\label{0407im}
\end{figure}

\begin{figure}
	\includegraphics[width=\columnwidth]{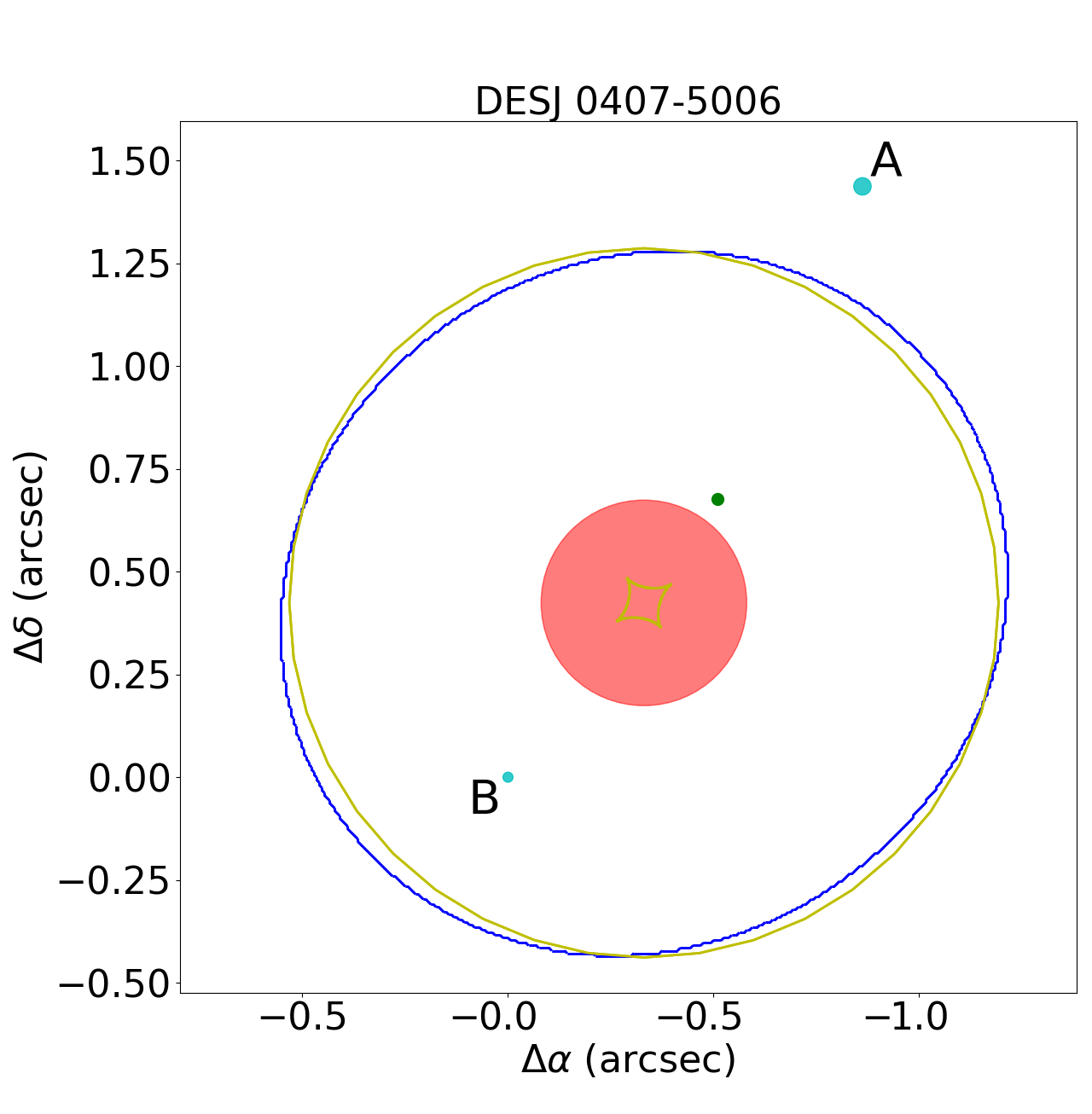}
	\caption{DESJ 0407-5006  lens model. Caustics are shown in yellow and the critical curve in blue. The source quasar is shown in green and the lens galaxy in red.}
	\label{mod0407}
\end{figure}

\begin{table*}
	\caption{Observed and best fit modeled parameters for DESJ 0407-5006. Astrometry, flux ratios and time delays with respect to image B. Since we do not have the redshift of the lensing galaxy, time delays are scaled  such that $\Delta_t=\Delta_t^\star (1+z_l) \frac{D_L}{D_{LS}}$ days.} 
	\label{0407ast}

	\begin{center}
		\begin{tabular}{c|c|c|c|c|c|c|c|c|c|c|c|c|c}
			\hline
			Component &   \multicolumn{3}{c}{Astrometry}   &  & \multicolumn{1}{c}{Flux Ratio} &  &                              \multicolumn{7}{c}{Model output parameters}                               \\ \hline
			          & $\Delta$ RA & $\Delta$ DEC & error &  &               i                &  & $\Delta$ RA & $\Delta$ DEC & $\kappa$ & $\gamma$ & $\mu$ &  FR  & \multicolumn{1}{c}{$\Delta t^\star$} \\ \hline
			    A     &   -0.863    &    1.437     & 0.002 &  &              3.17              &  &   -0.862    &    1.437     &   0.38   &   0.40   & 4.48  & 3.00 &                -25.46                \\
			    B     &     0.0     &     0.0      & 0.004 &  &              1.0               &  &     0.0     &     0.0      &   0.81   &   0.84   & -1.50 & 1.00 &                 0.0                  \\
			    G     &   -0.331    &    0.424     & 0.06  &  &                                &  &   -0.331    &    0.424     &          &          &       &      &                                      \\ \hline
		\end{tabular}
	\end{center}
\end{table*}

\subsection{Contaminants}

\begin{figure*}
	\centering
	\begin{subfigure}[b]{0.475\textwidth}
		\centering
		\includegraphics[width=\columnwidth]{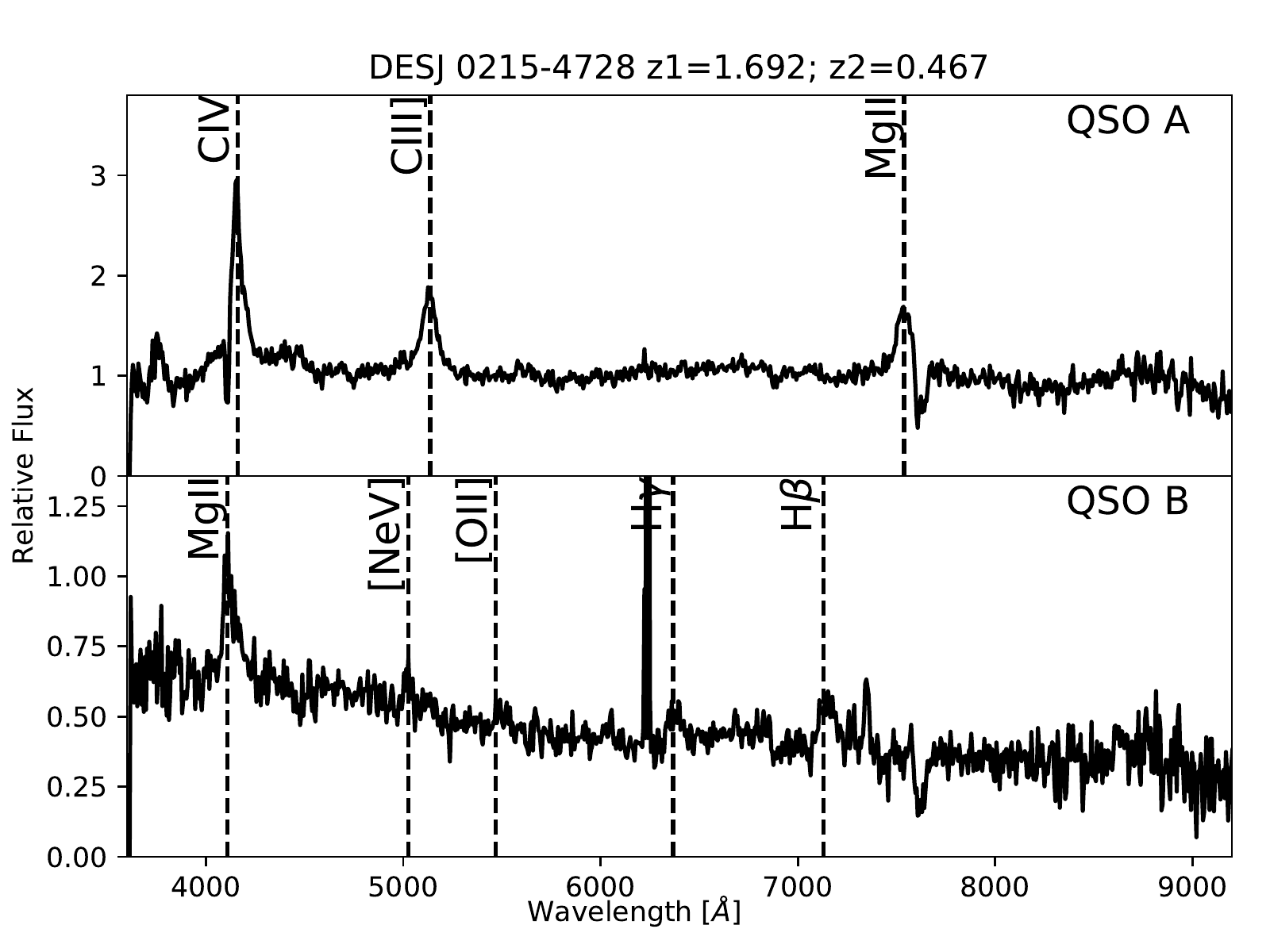}
		\caption{DESJ 0215-4728}
	\end{subfigure}%
	~ 
	\begin{subfigure}[b]{0.475\textwidth}
		\centering
		\includegraphics[width=\columnwidth]{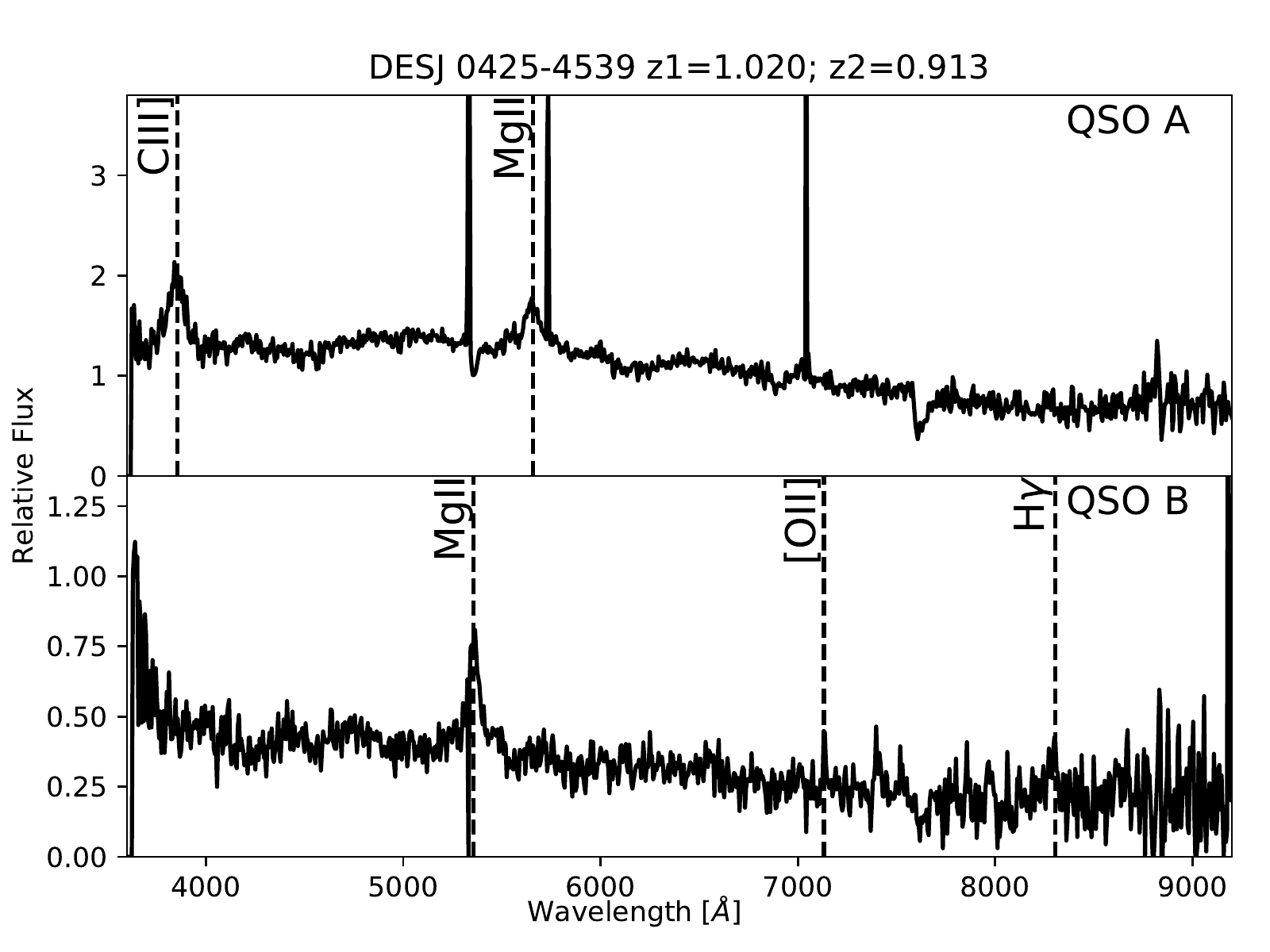}
		\caption{DESJ 0425-4539}
	\end{subfigure}
	~ 
	\begin{subfigure}[b]{0.475\textwidth}
		\centering
		\includegraphics[width=\columnwidth]{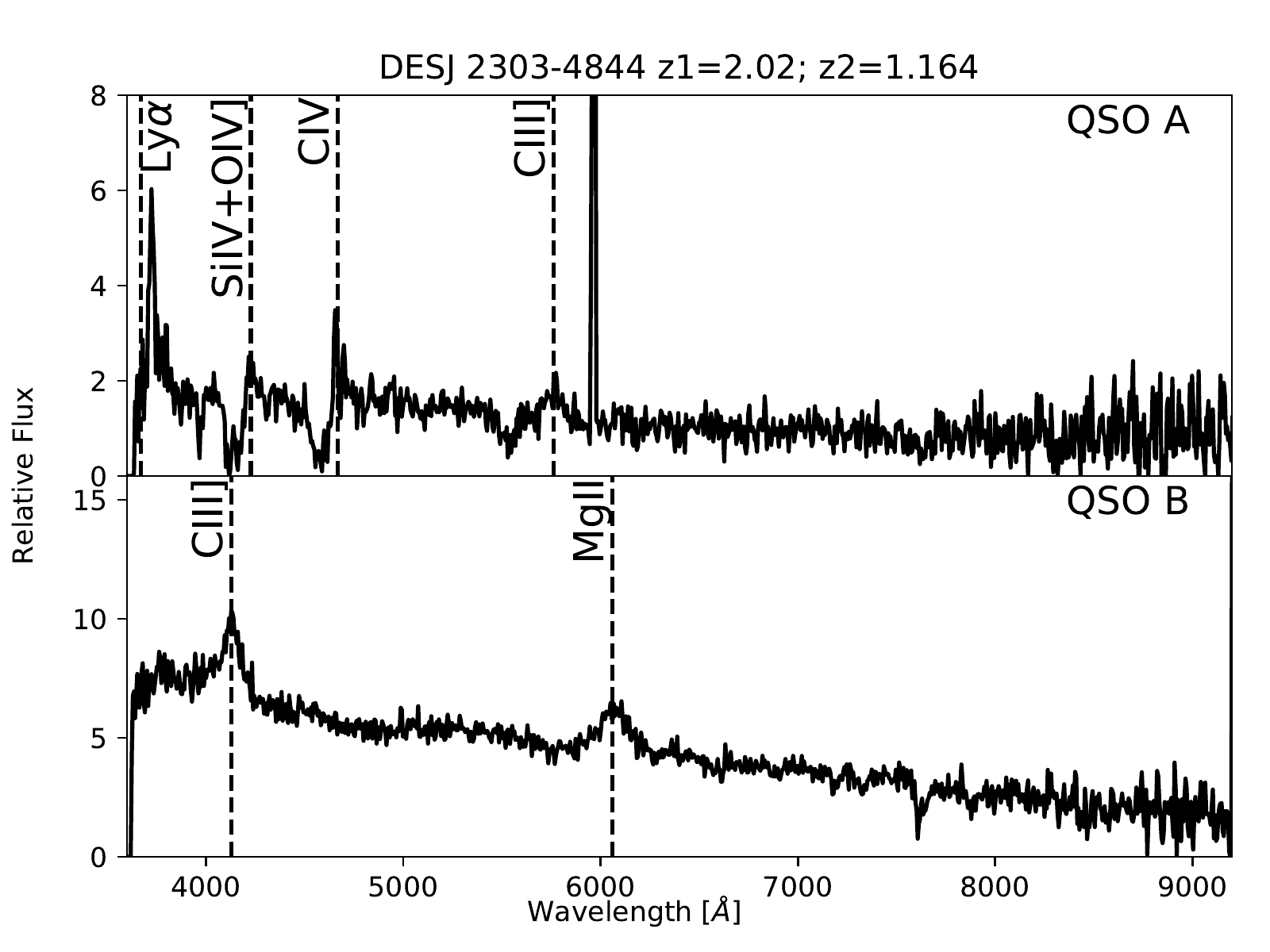}
		\caption{DESJ 2303-4844}
	\end{subfigure}
	~ 
	\begin{subfigure}[b]{0.475\textwidth}
		\centering
		\includegraphics[width=\columnwidth]{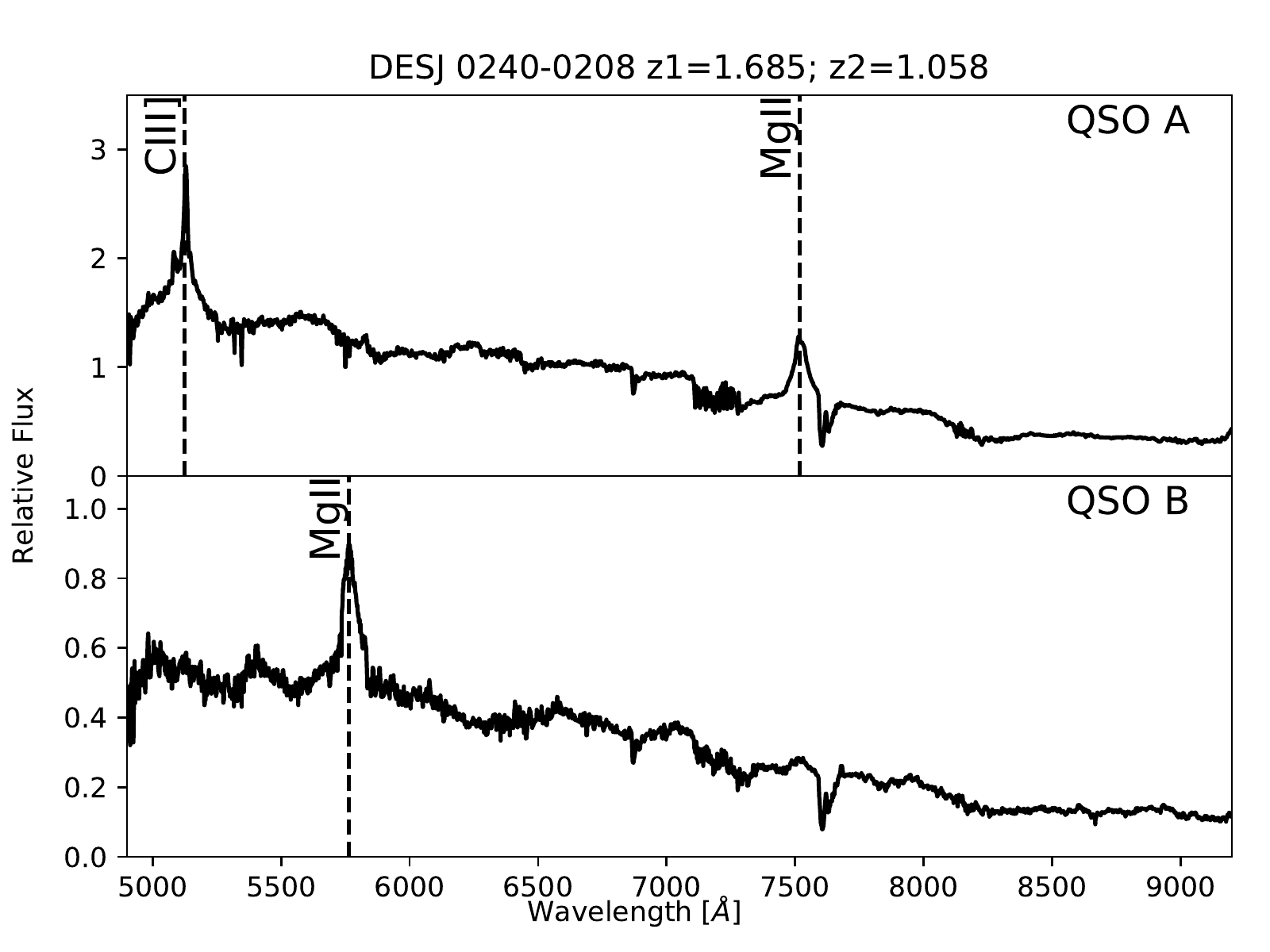}
		\caption{DESJ 0240-0208}
	\end{subfigure}
	\caption{Projected binaries spectra. Segmented lines show the identified emission lines used to measure the redshifts shown above each panel.}
	\label{fig:rojbin}
\end{figure*}

Several interlopers have been identified from our spectroscopic follow up. These are all listed in Table \ref{tab:Conc}. In particular, four projected double quasars are identified. Their spectra with line identification is shown in Fig. \ref{fig:rojbin}. Most of the confirmed contaminants (15) include at least one quasar (4 quasar pairs, 4 quasar-galaxy pairs and 7 quasar-star pairs), with the eight remaining being galaxies: 6 with measured strong narrow emission lines and 2 due to their extended shape from high resolution SOAR SAMI imaging. 

\label{sec:binaries}

\section{Inconclusive Systems}

Despite the follow up campaign, the nature of 9 out of the 34 systems still remains inconclusive. However, two subcategories of inconclusive systems are identified. A summary of their properties and follow up data is shown in Table \ref{tab:Conc}. 

\subsection{Nearly Identical Quasar Pairs (NIQs)}
\label{sec:NIPs}

``Nearly Identical Quasar Pairs'' (NIQs) are the systems followed up where we have obtained two resolved nearly identical quasar spectra of candidates, but we have not been able to identify a lensing galaxy in imaging. Seven of the systems in our follow up sample, are classified as NIQs. Most of these systems should be considered as very likely lensed quasars but we currently lack the imaging and/or spectroscopic observations to confirm them as such. The spectra of the image pairs in this category are shown in Fig. \ref{fig:NIPs}. Table \ref{tab:niqmag} shows a crude estimation of the minimum brightness of the unidentified lens galaxy between the candidate lensed images using the image separation and source redshift \citep[see description of the method in][]{treu2018}. Even though this is only an estimation, we note that several lens galaxies should have been bright enough for detection in the survey imaging. Furthermore, out of the seven systems in the NIQ category, two show spectroscopic evidence supporting the interpretation of two different quasars at identical redshifts (i.e. physical binaries) as discussed below.  However, conservatively and due to lack of evidence we still classify them as nearly identical pairs.

\begin{table}
	
	\caption{Maximum magnitude in the SDSS g, r, i and z band for the possible lenses between the NIQs, obtained from the lensing mass expected due to the image separation of the candidate lensed images and their measured redshifts.}
	\label{tab:niqmag} 
	\begin{center}
		\begin{tabular}{lllll}
 \hline
 ID                   &g	 &r &i &z \\ \hline
 DESJ 005817.0-612004 &24.6 & 21.9&20.6 &	19.6 \\
 DESJ 012006.4-435441 &28.7 & 26.2&24.6 &	23.3 \\
 DESJ 054430.6-592237 &26.5 & 23.8&22.5 &	21.6 \\
 DESJ 214148.9-462946 &25.9 & 23.1&21.8 &	20.5 \\
 DESJ 220039.1-471900 &25.1 & 22.2&21.0 &   19.7 \\
 DESJ 225007.9-604724 &24.3 & 22.1&20.7 &	20.2 \\
 DESJ 233713.7+005611 &23.2 & 21.4&20.3 &	19.7 \\ \hline
 		\end{tabular} 
	\end{center} 
\end{table}

\subsubsection{DESJ 0120-4354}

This system at z$\approx$1.91 shows a very broad double absorption blue-ward of the CIV line. At first sight this might be attributable to a lower redshift intervening system in between the two quasar images. However, the strength and broadness of the lines is not consistent with any known doublets. Another explored possibility was that they are two strong intervening MgII systems, however due to the small separation of the two quasar images (0\farcs84) these systems would appear in quasar image B. Furthermore, the profile of the SiIV+OIV] lines shows a difference that could not be attributed to microlensing \citep[see e.g.][]{sluse2012}. Finally, image B shows a slightly lower redshift at z=1.909 (versus image A at z=1.911). As such, both quasar images could indeed be different systems, one of them a double BAL \citep{korista1993}. This system has also been independently discovered by \cite{ostrovski2018}.

\subsubsection{DESJ 2141-4629}

Close inspection of the CIV emission line reveals very strong differences. In particular, the peak of the CIV emission in each image is red and blue shifted respectively. Furthermore, each peak coincides with an absorption in the other \citep[see e.g.][]{hennawi2006}. Two absorption systems have been identified at z=0.711 and z=1.420 from FeII triplet ($\sim$2370\AA{}), FeII doublet ($\sim$2600\AA{}) and MgII doublet ($\sim$2800\AA{}) absorption. Given our current interpretation of the quasar spectra along the fact that the absorption lines resulting from these systems are too narrow to be due to a massive galaxy, we do not believe these are indicative of a lensing galaxy \citep[see, e.g., the discussion of broadness of absorption lines due to intervening systems in lensed quasars in][]{auger2008}.

\begin{figure*}
	\centering
	\begin{subfigure}[b]{0.475\textwidth}
		\centering
		\includegraphics[width=\columnwidth]{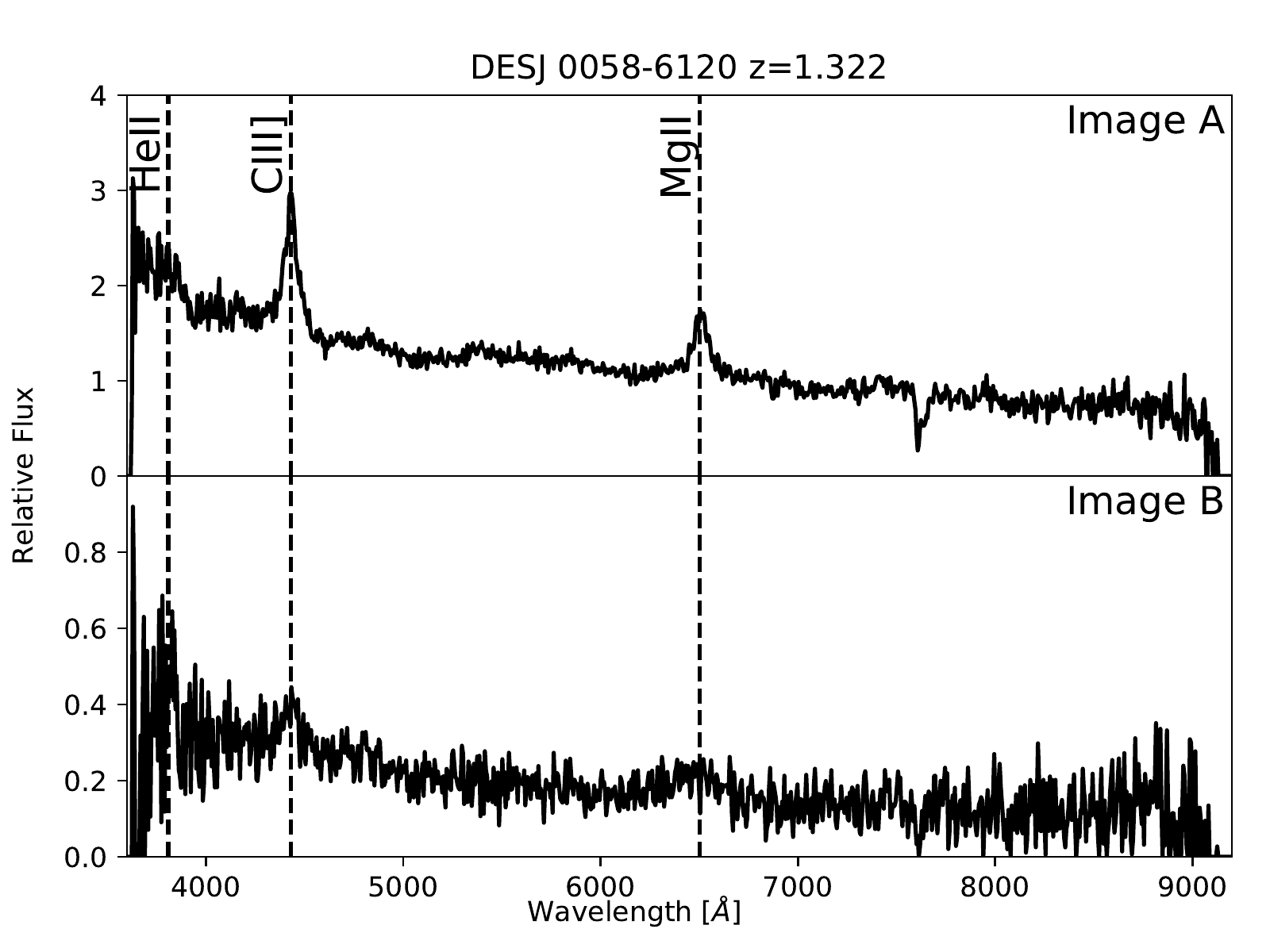}
		\caption{DESJ 0058-6120}
	\end{subfigure}%
	~
	\begin{subfigure}[b]{0.475\textwidth}
	\centering
	\includegraphics[width=\columnwidth]{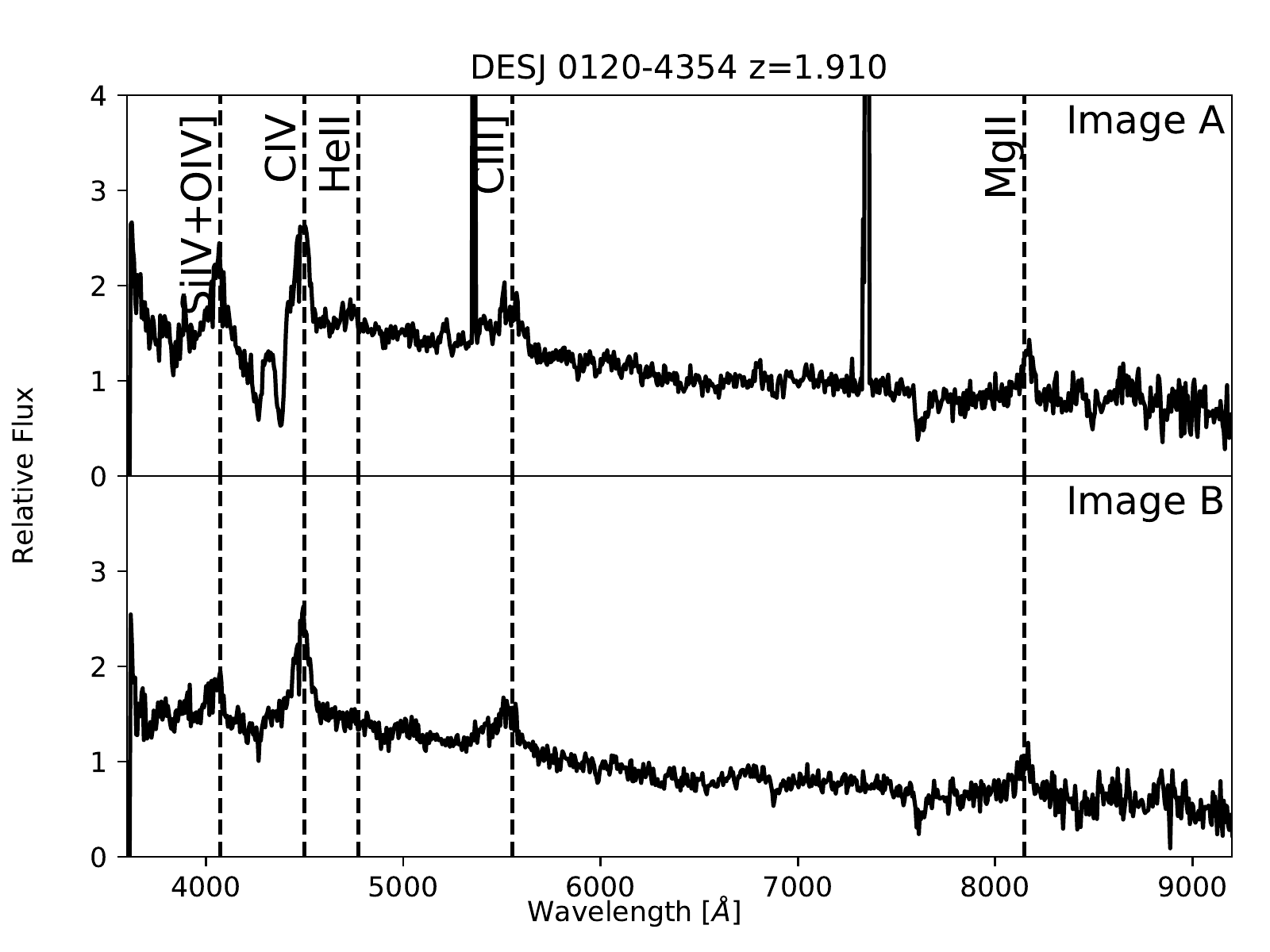}
	\caption{DESJ 0120-4354}
	\end{subfigure}%

	~ 
	\begin{subfigure}[b]{0.475\textwidth}
	\centering
	\includegraphics[width=\columnwidth]{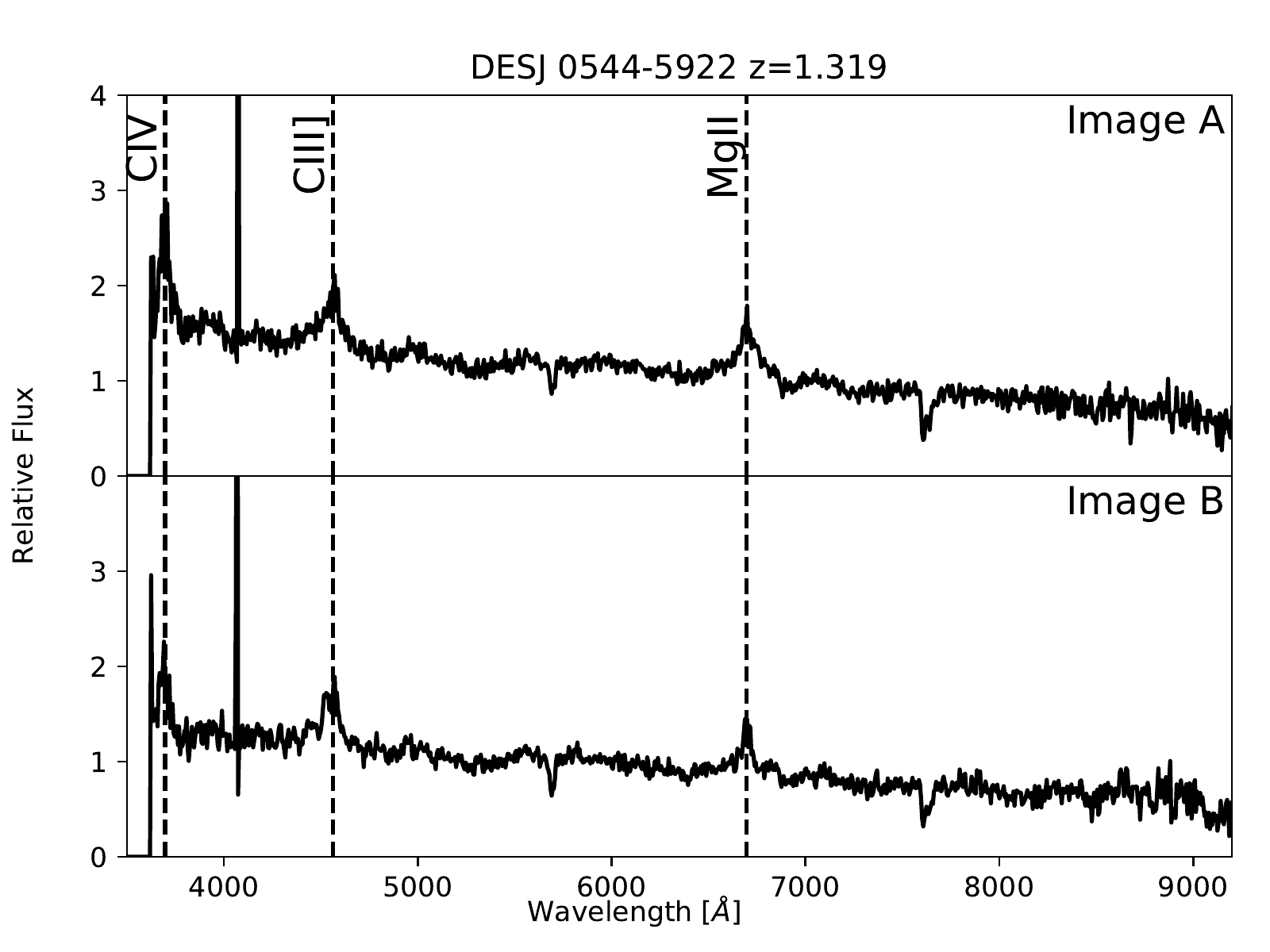}
	\caption{DESJ 0544-5922}
	\end{subfigure}
	~ 
	\begin{subfigure}[b]{0.475\textwidth}
	\centering
	\includegraphics[width=\columnwidth]{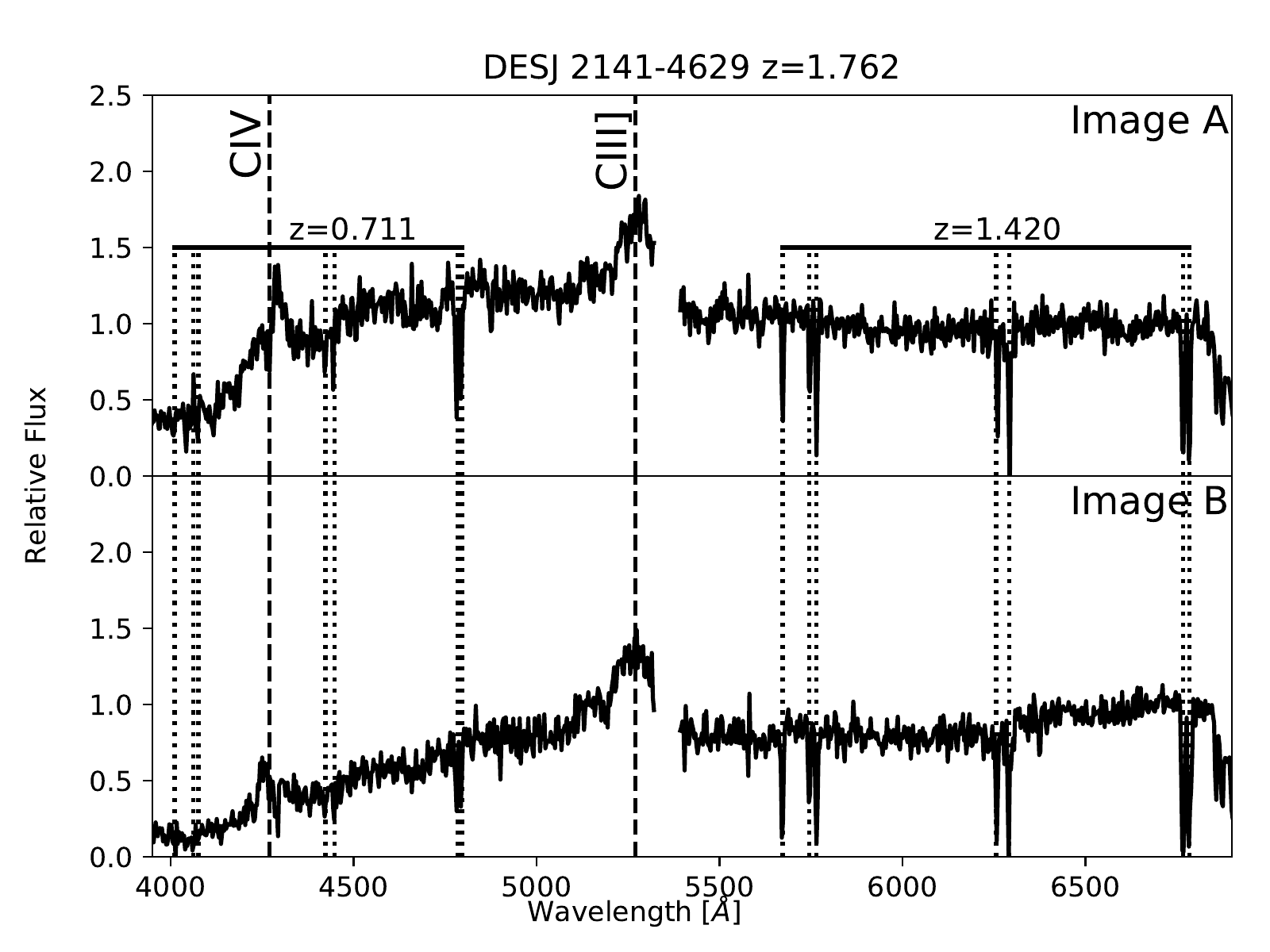}
	\caption{DESJ 2141-4629}
	\end{subfigure}
	~	
	\begin{subfigure}[b]{0.475\textwidth}
		\centering
		\includegraphics[width=\columnwidth]{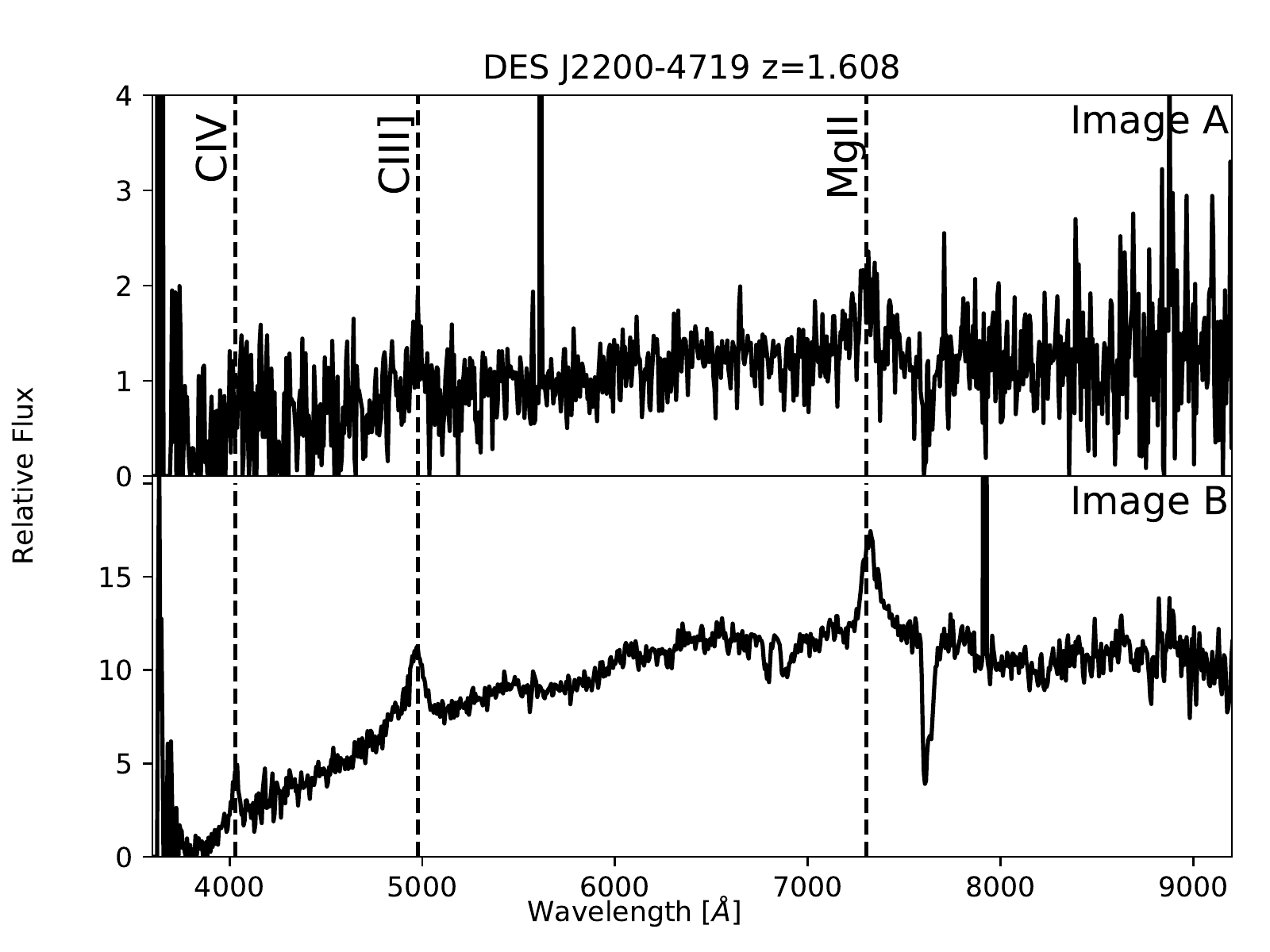}
		\caption{DESJ 2200-4719}
	\end{subfigure}
	~ 
	\begin{subfigure}[b]{0.475\textwidth}
		\centering
		\includegraphics[width=\columnwidth]{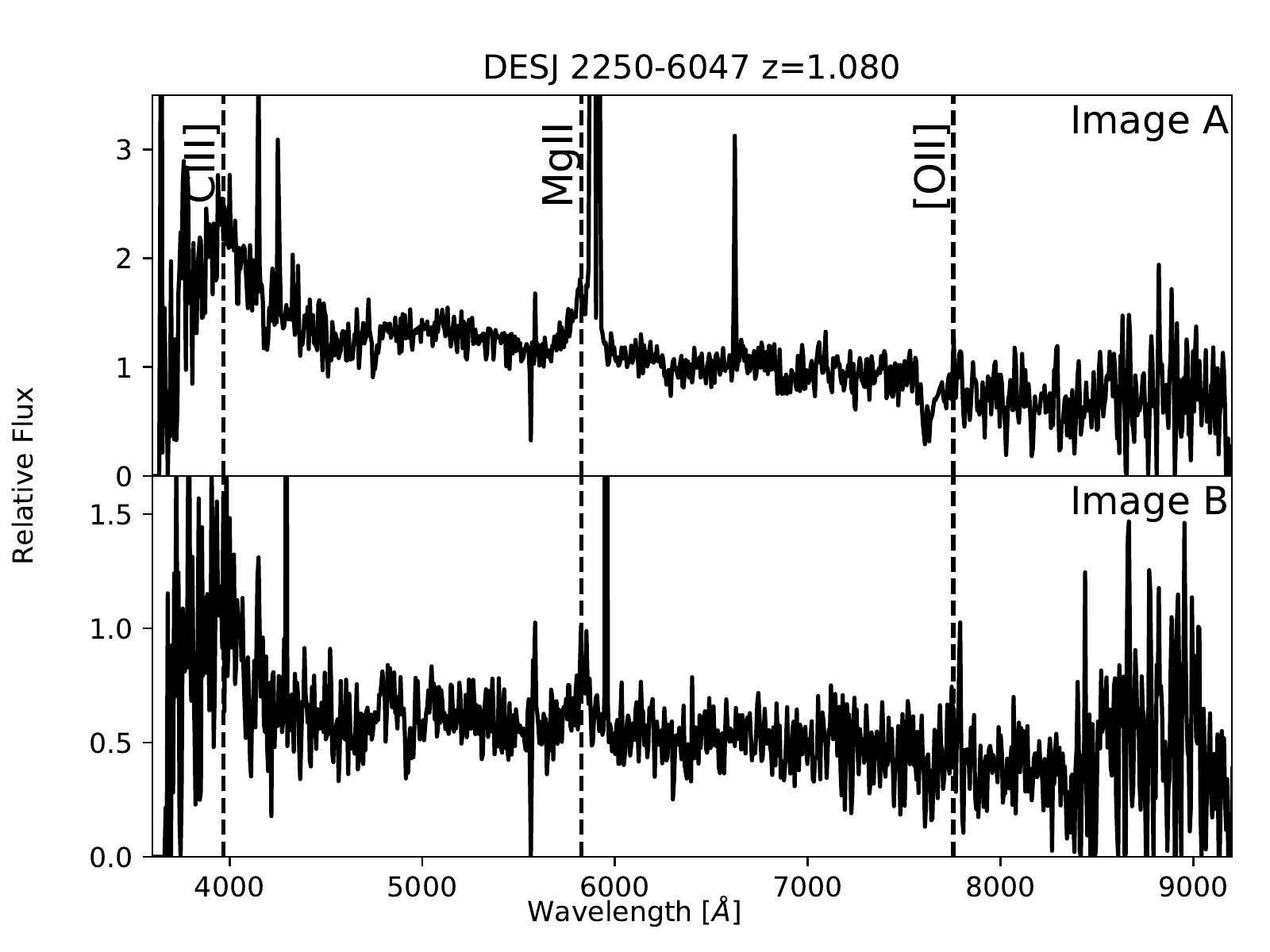}
		\caption{DESJ 2250-6047}
	\end{subfigure}
	\caption{Nearly Identical Quasar Pairs (NIQs) spectra. Segmented lines show the identified emission lines used to measure the redshifts shown above each panel.}
    \label{fig:NIPs}
\end{figure*}
\begin{figure*}\ContinuedFloat
	\centering	
	\begin{subfigure}[b]{0.475\textwidth}
		\centering
		\includegraphics[width=\columnwidth]{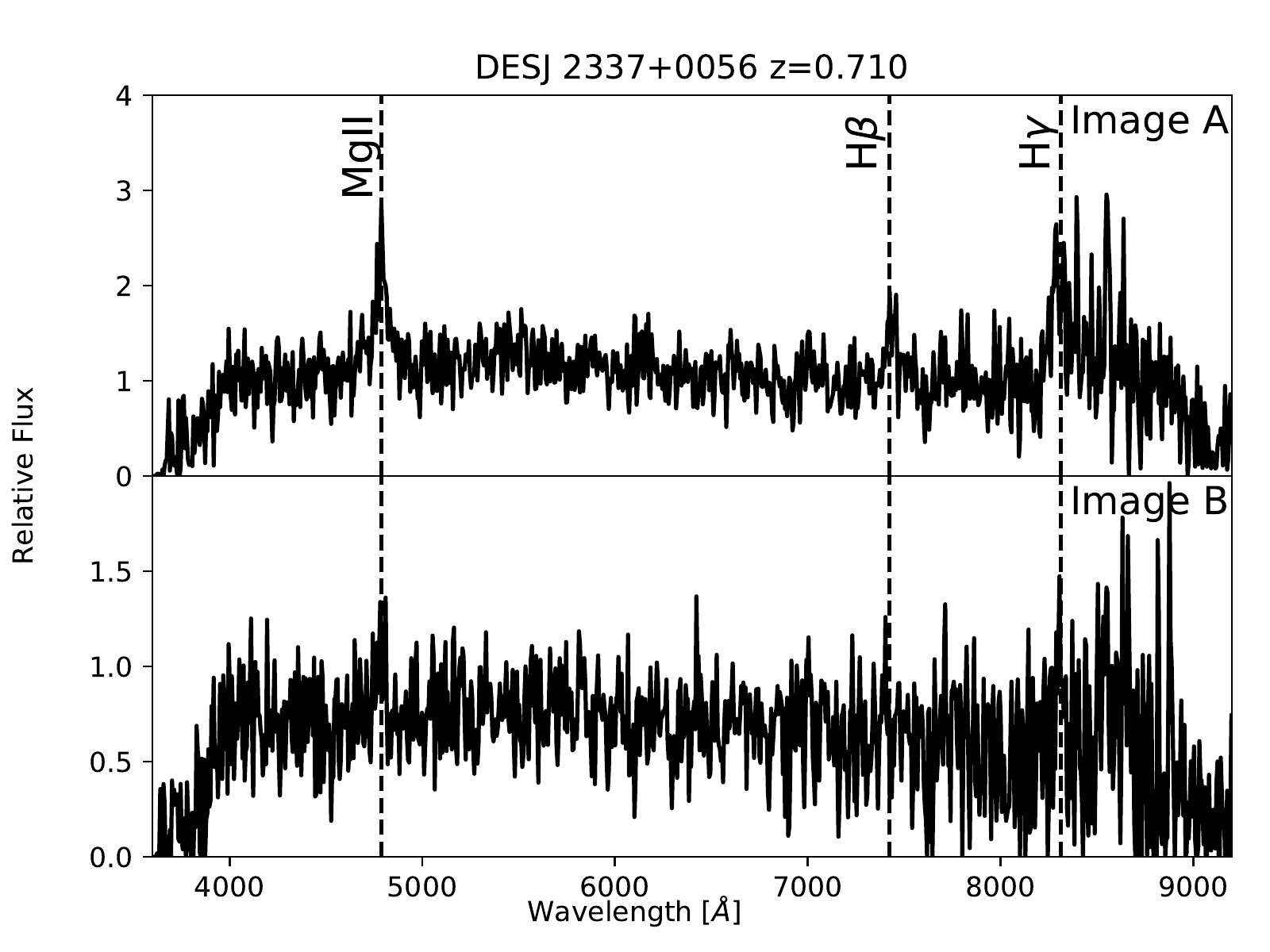}
		\caption{DESJ 2337+0056}
	\end{subfigure}	
	\caption{\textit{continued} Nearly Identical Quasar Pairs (NIQs) spectra. Segmented lines show the identified emission lines used to measure the redshifts shown above each panel. \textit{continued}}
	\label{fig:NIPs}
\end{figure*}

\subsection{Otherwise inconclusive}

Two additional systems followed up contain at least one image that has been confirmed to have a quasar spectrum. The spectroscopic flux obtained for the fainter image does not allow to identify its nature and as such also remain inconclusive.

\section{Summary and Conclusions}
	
We have followed up 34 lensed quasar candidate systems selected by a mixture of multi-band psf fitting and color selection in the fields of the Dark Energy Survey. From this follow up, two systems have conclusively been classified as gravitationally lensed quasars (one quadruple and one double). Seven systems are likely double lensed quasars based on almost indistinguishable spectra between the two candidate images, but a hitherto unidentified lensing galaxy between them (Nearly Identical Quasar Pairs or NIQs). Two systems remain inconclusive since only one spectrum of the pairs is identified as a quasar. The remaining 23 systems have been confirmed as non lensed systems (15 are indeed quasars with non identical companions and 8 are pairs of compact low redshift emission regions). Thus, removing the two completely inconclusive systems, we see our selection has a effectiveness of 30\% selecting identical pairs of quasars, or 50\% considering pairs of quasars (since 4 of the non-lensed systems are projected double quasar systems).

The two lensed quasar systems have been identified thanks to deeper higher resolution images where a lensing galaxy could be clearly identified, even when no absorption lines from them have been identified in the spectra. Deeper and higher resolution spectroscopy, will allow us to measure their redshifts. Nevertheless, simple lens models were performed for both systems, and they reveal that a very minor quadrupole (internal or external) is required in order to reproduce the positional constraints, consistent with the observed light profile of the lensing galaxy. For the quadruply imaged quasar DES J0405-3308 two images show mild flux ratio inconsistencies with respect to the lens model, which could be consistent with microlensing or dust extinction from the lens galaxy. The same is true to an event lesser extent for DES J0407-5006. 

We have estimated the minimum brightness in the griz SDSS photometric bands for the possible lensing galaxy in the seven NIQs. Our estimations show that, if indeed lensed quasars, at least three of them should have been detected in the survey imaging image subtraction (i<21.0). As such, priorities for follow up should include a prior on image separation (larger separations imply brighter lenses, and thus lower probability of being lensed quasars if not detected in the survey imaging). We do note, however, that our magnitude estimates depend on several assumptions including lens redshifts. Furthermore, lens galaxies could well be hiding under the PSF wings of the, sometimes, much brighter quasar images, rendering the nominal depth optimistic. Nonetheless, some of these systems are still worthy of follow up to confirm their nature.

\section*{Acknowledgements}

 T. A. acknowledges support by proyecto FONDECYT 11130630 and by the Ministry for the Economy, Development, and Tourism's Programa Inicativa Cient\'{i}fica Milenio through grant IC 12009, awarded to The Millennium Institute of Astrophysics (MAS). T.T. and V.M. acknowledge support by the
 Packard Foundation through a Packard Research Fellowship to T.T. T.T.
 acknowledges support by the National Science Foundation through grant
 AST-1450141.
 
Funding for the DES Projects has been provided by the U.S. Department of Energy, the U.S. National Science Foundation, the Ministry of Science and Education of Spain, 
the Science and Technology Facilities Council of the United Kingdom, the Higher Education Funding Council for England, the National Center for Supercomputing 
Applications at the University of Illinois at Urbana-Champaign, the Kavli Institute of Cosmological Physics at the University of Chicago, 
the Center for Cosmology and Astro-Particle Physics at the Ohio State University,
the Mitchell Institute for Fundamental Physics and Astronomy at Texas A\&M University, Financiadora de Estudos e Projetos, 
Funda{\c c}{\~a}o Carlos Chagas Filho de Amparo {\`a} Pesquisa do Estado do Rio de Janeiro, Conselho Nacional de Desenvolvimento Cient{\'i}fico e Tecnol{\'o}gico and 
the Minist{\'e}rio da Ci{\^e}ncia, Tecnologia e Inova{\c c}{\~a}o, the Deutsche Forschungsgemeinschaft and the Collaborating Institutions in the Dark Energy Survey. 

The Collaborating Institutions are Argonne National Laboratory, the University of California at Santa Cruz, the University of Cambridge, Centro de Investigaciones Energ{\'e}ticas, 
Medioambientales y Tecnol{\'o}gicas-Madrid, the University of Chicago, University College London, the DES-Brazil Consortium, the University of Edinburgh, 
the Eidgen{\"o}ssische Technische Hochschule (ETH) Z{\"u}rich, 
Fermi National Accelerator Laboratory, the University of Illinois at Urbana-Champaign, the Institut de Ci{\`e}ncies de l'Espai (IEEC/CSIC), 
the Institut de F{\'i}sica d'Altes Energies, Lawrence Berkeley National Laboratory, the Ludwig-Maximilians Universit{\"a}t M{\"u}nchen and the associated Excellence Cluster Universe, 
the University of Michigan, the National Optical Astronomy Observatory, the University of Nottingham, The Ohio State University, the University of Pennsylvania, the University of Portsmouth, 
SLAC National Accelerator Laboratory, Stanford University, the University of Sussex, Texas A\&M University, and the OzDES Membership Consortium.

Based in part on observations at Cerro Tololo Inter-American Observatory, National Optical Astronomy Observatory, which is operated by the Association of 
Universities for Research in Astronomy (AURA) under a cooperative agreement with the National Science Foundation.

The DES data management system is supported by the National Science Foundation under Grant Numbers AST-1138766 and AST-1536171.
The DES participants from Spanish institutions are partially supported by MINECO under grants AYA2015-71825, ESP2015-88861, FPA2015-68048, SEV-2012-0234, SEV-2016-0597, and MDM-2015-0509, 
some of which include ERDF funds from the European Union. IFAE is partially funded by the CERCA program of the Generalitat de Catalunya.
Research leading to these results has received funding from the European Research
Council under the European Union's Seventh Framework Program (FP7/2007-2013) including ERC grant agreements 240672, 291329, and 306478.
We  acknowledge support from the Australian Research Council Centre of Excellence for All-sky Astrophysics (CAASTRO), through project number CE110001020.

This manuscript has been authored by Fermi Research Alliance, LLC under Contract No. DE-AC02-07CH11359 with the U.S. Department of Energy, Office of Science, Office of High Energy Physics. The United States Government retains and the publisher, by accepting the article for publication, acknowledges that the United States Government retains a non-exclusive, paid-up, irrevocable, world-wide license to publish or reproduce the published form of this manuscript, or allow others to do so, for United States Government purposes.

(Some of) The data presented herein were obtained at the W. M. Keck
Observatory, which is operated as a scientific partnership among the
California Institute of Technology, the University of California and
the National Aeronautics and Space Administration. The Observatory was
made possible by the generous financial support of the W. M. Keck
Foundation. The authors wish to recognize and acknowledge the very
significant cultural role and reverence that the summit of Maunakea
has always had within the indigenous Hawaiian community.  We are most
fortunate to have the opportunity to conduct observations from this
mountain.  Based in part on observations obtained at the Southern
Astrophysical Research (SOAR) telescope, which is a joint project of
the Minist\'{e}rio da Ci\^{e}ncia, Tecnologia, Inova\c{c}\~{a}os e
Comunica\c{c}\~{a}oes (MCTIC) do Brasil, the U.S. National Optical
Astronomy Observatory (NOAO), the University of North Carolina at
Chapel Hill (UNC), and Michigan State University (MSU). Based in part on observations made with ESO Telescopes at the La Silla
Paranal Observatory. This publication makes use of data products from
the Wide-field Infrared Survey Explorer, which is a joint project of
the University of California, Los Angeles, and the Jet Propulsion
Laboratory/California Institute of Technology, funded by the National
Aeronautics and Space Administration. This publication includes data gathered with the 6.5 meter Magellan Telescopes located at Las Campanas Observatory, Chile.




\bibliographystyle{mnras}
\bibliography{STRIDES} 





\bsp	
\label{lastpage}
\end{document}